\documentclass[desactivate]{aa} 

\usepackage{graphicx}
\usepackage{subfig}
\usepackage{txfonts}
\usepackage{csvsimple,booktabs}
\usepackage{filecontents}

\usepackage{pdfpages}

\usepackage{natbib}
\usepackage[colorlinks=true,allcolors=blue]{hyperref} 
\bibpunct{(}{)}{;}{a}{}{,}             
\begin{document}

   \title{Breaking degeneracies in exoplanetary parameters through self-consistent atmosphere--interior modelling}

   \author{C. Wilkinson\inst{1}        
          \and B. Charnay\inst{1}
          \and S. Mazevet\inst{2}
          \and A.-M. Lagrange\inst{1}
          \and A. Chomez\inst{1,3}
          \and V. Squicciarini\inst{1,4}
          \and E. Panek\inst{5}
          \and J. Mazoyer\inst{1}
          }

   \institute{
        LESIA, Observatoire de Paris, Universit\'{e} PSL, CNRS, 5 Place Jules Janssen, 92190 Meudon, France \\
        \email{christian.wilkinson@obspm.fr}
        \and
        Observatoire de la Côte d'Azur, Université Côte d'Azur, 96. Boulevard de l'observatoire 06300 Nice, France
        \and
        Univ. Grenoble Alpes, CNRS-INSU, Institut de Planétologie et d'Astrophysique de Grenoble (IPAG) UMR 5274, Grenoble, F-38041, France
        \and
        INAF -- Osservatorio Astronomico di Padova, Vicolo dell'Osservatorio 5, Padova, Italy, I-35122
        \and
        Institut d’Astrophysique de Paris (CNRS, Sorbonne Université), 98bis Bd. Arago, 75014 Paris, France
    }


    \abstract
    {With a new generation of observational instruments largely dedicated to exoplanets (i.e. JWST, ELTs, PLATO, and Ariel) providing atmospheric spectra and mass and radius measurements for large exoplanet populations, the planetary models used to understand the findings are being put to the test.}
    {We seek to develop a new planetary model, the Heat Atmosphere Density Evolution Solver (HADES), which is the product of self-consistently coupling an atmosphere model and an interior model, and aim to compare its results to currently available findings.}
    {We conducted atmospheric calculations under radiative-convective equilibrium, while the interior is based on the most recent and validated ab initio equations of state. We pay particular attention to the atmosphere--interior link by ensuring a continuous thermal, gravity, and molecular mass profile between the two models.}
    {We applied the model to the database of currently known exoplanets to characterise intrinsic thermal properties. In contrast to previous findings, we show that intrinsic temperatures (T$_{int}$) of 200-400 K ---increasing with equilibrium temperature--- are required to explain the observed radius inflation of hot Jupiters. In addition, we applied our model to perform `atmosphere--interior' retrievals by Bayesian inference using observed spectra and measured parameters. This allows us to showcase the model using example applications, namely to WASP-39 b and 51 Eridani b. For the former, we show how the use of spectroscopic measurements can break degeneracies in the atmospheric metallicity (Z) and intrinsic temperature. We derive relatively high values of Z = 14.79$_{-1.91}^{+1.80}\times$solar and T$_{int}=297.39_{-16.9}^{+8.95}$K, which are necessary to explain the radius inflation and the chemical composition of WASP-39 b. With this example, we show the importance of using a self-consistent model with the radius being a constrained parameter of the model and of using the age of the host star to break radius and mass degeneracies. When applying our model to 51 Eridani b, we derive a planet mass M$_p=3.13_{-0.04}^{0.05}$ M$_{J}$ and a core mass M$_{core}=31.86_{+0.32}^{-0.18}$ M$_{E}$, suggesting a potential formation by core accretion combined with a `hot start' scenario.}
    {We conclude that self-consistent atmosphere--interior models efficiently break degeneracies in the structure of both transiting and directly imaged exoplanets. Such tools have great potential to interpret current and future observations, thereby providing new insights into the formation and evolution of exoplanets.}


\titlerunning{Self-consistent atmosphere-interior modelling for exoplanets}

\maketitle
%

\section{Introduction}

In recent decades, the discovery of strongly irradiated giant exoplanets \citep{mayor_jupiter-mass_1995} has opened new avenues of exploration in planetary science. These exoplanets, named hot Jupiters, exhibit strikingly different atmospheres compared to the giant planets in our Solar System 
\citep[e.g.,][]{guillot_giant_1996, guillot_evolution_2002}, rendering them compelling subjects for in-depth investigation. Understanding their unique atmospheric characteristics and their impact on planetary observables has become crucial for advancing our knowledge of exoplanetary systems. Furthermore, the discovery of directly imageable exoplanets \citep[e.g.][]{chauvin_giant_2004, chauvin_companion_2005, marois_direct_2008, marois_images_2010,  lagrange_probable_2009} enables us to explore the early, still-warm stages of giant planets, providing valuable insights into their formation processes.

A better understanding of the gas giants of our Solar System has had meaningful consequences when it comes to studying exoplanets. On one hand, the difficulty in accurately evaluating the evolution of the ice giants raises questions regarding their internal structure \citep{scheibe_thermal_2019}. On the other, the results from the Juno mission highlight the ongoing difficulty in understanding the layout of Jupiter's planetary core, combined with the limitations of current equations of state \citep[e.g.][]{nettelmann_theory_2021, idini_gravitational_2022, militzer_juno_2022-1, howard_jupiters_2023}. We still have far more data for our close neighbours than we do for exoplanets, which makes the former ideal objects of study to fine tune current modelling tools. However, the growing number of discovered exoplanets represents an ideal test bed for evaluating and exploring the wider uses of these tools and for performing statistical analyses.

In addressing these complexities, the integration of comprehensive grids of model atmospheres has played a pivotal role \citep[e.g.][]{baraffe_evolutionary_2003, phillips_new_2020, marley_sonora_2021}. Each successive generation of models has been designed to refine and augment various physical aspects, whether through improved interior equations of state (EOS) and disequilibrium chemistry \citep{phillips_new_2020} or the exploration of the effects of varying C/O ratios \citep{marley_sonora_2021}. As emphasised by \cite{marley_sonora_2021}, the diversity of model grids is essential for their complementary nature and intercomparison, enhancing their overall efficacy. Furthering the development of exoplanet model grids requires the inclusion of stellar irradiation and improvement of the coherence between planet interior and atmosphere. The present work addresses both issues.

The modelling of exoplanet interiors often faces significant degeneracy, underscoring the importance of integrating interior models with atmospheric properties. This integration is exemplified in \cite{bloot_exoplanet_2023}, where atmospheric metallicity is used to constrain interior designs. Additionally, \cite{muller_synthetic_2021} explore how atmospheric properties contribute to constraining planetary bulk composition. Their study of 51 Eri b showcases the use of observed luminosity, derived metallicity, and synthetic evolution tracks to estimate mass and bulk metallicity. The strength of such an approach lies in its coherent connection of various parameters. In our work, we propose to advance this approach further by integrating a more sophisticated atmospheric model capable of generating synthetic spectra. This advancement grants us the ability to derive atmospheric properties from observations while maintaining constraints imposed by the interior. \cite{guzman-mesa_chemical_2022} exemplify the strengths of this type of model when extracting the properties of GJ436 b, a sub-Neptune-like planet, through a coupled interior--atmosphere model with spectral retrieval. Our understand- ing of Sub-Neptunes is plagued by degeneracy, as highlighted by \cite{valencia_bulk_2013}. Expanding upon this framework, we extend the approach to encompass Jupiter and super-Jupiter-like planets for both transit and direct imaging studies.

When irradiated, predicting the observed radii of hot Jupiters  is distinctly challenging due to the inflated nature \citep{fortney_hot_2021}. Accurately modelling these bodies necessitates the consideration of stronger intrinsic thermal fluxes than expected for the ages of the systems they are found in. Properly accounting for this phenomenon is crucial not only for 1D model grids but also for 3D general circulation models (GCMs). Recent studies have demonstrated the possibility of establishing a connection between the equilibrium temperature of a hot Jupiter and its intrinsic thermal flux, providing a plausible explanation for the observed radius \citep{thorngren_bayesian_2018, thorngren_intrinsic_2019}. Currently, the two favoured hypotheses are ohmic dissipation \citep{batygin_inflating_2010} and advection of heat by atmospheric circulation \citep{tremblin_advection_2017}. With the present work, we aim to address this issue by directly coupling atmospheric and interior modellings so as to evaluate the evolution of irradiated planets by resolving the thermal evolution equation across a grid of planets in equilibrium at different intrinsic temperatures. 

In this work, we try to add to the models available by producing a self-consistent interior-atmosphere model that ties physical parameters together. We believe that this will allow better constraint of parameters such as mass, intrinsic temperature, metallicity, core mass, and radius from observations. In section \ref{Model description}, we explore the atmosphere and interior models used and how we link the two into one that is self-consistent in order to build grids of models. We make literature comparisons to verify the veracity of the results. In section \ref{Stellar irradiation}, we evaluate how exoplanets are affected by stellar irradiation. In Section \ref{Breaking degeneracies}, we discuss the various planetary parameters and how they can lead to degeneracies. In section \ref{Grid interpolation and retrieval}, we explore how Bayesian inference tools can be used with our model to infer parameters that explain an observed spectrum. Finally, in Sections \ref{Case study : WASP-39 b} and \ref{Case study : 51 Eridani b}, we show how the model can be applied to infer planetary parameters from observations.

\section{Model description}\label{Model description}
\subsection{Atmosphere model}\label{Atmosphere model}
The atmosphere model used in this work is \verb+Exo-REM+ \citep{baudino_interpreting_2015, baudino_toward_2017, charnay_self-consistent_2018, blain_1d_2021}. \verb+Exo-REM+ is a 1D radiative-equilibrium model that calculates fluxes using the two-stream approximation assuming hemispheric closure. Radiative-convective equilibrium is solved assuming that the net flux (radiative + convective) is conserved. The conservation of flux over the grid of pressure levels is solved iteratively using a constrained linear inversion method. Thermodynamic quantities in \verb+Exo-REM+ are linked to one another via the ideal gas law.
The vertical chemical profiles for the various species are calculated for a given temperature profile assuming some non-equilibrium chemistry between C-, O-, and N-bearing compounds. To do this, we use an analytical formulation based on a comparison of chemical time constants with vertical mixing time from \cite{zahnle_methane_2014}. For this calculation, vertical mixing is parameterised by an eddy mixing coefficient, K$_{zz}$. To maximise the comparability of the models, we chose to use a constant value for the K$_{zz}$, as done by \cite{phillips_new_2020}: we take $\log($K$_{zz})=10$ (with K$_{zz}$ in $\mathrm{cm^2/s}$). \cite{phillips_new_2020} demonstrate that the eddy mixing coefficient has limited effects on evolution models. We include the option to use equilibrium chemistry. Condensed species that would otherwise have formed clouds are removed.
\verb+Exo-REM+ is parameterised by a gravity at a pressure of 1 bar (g), an intrinsic temperature (T$_{int}$), an incoming stellar irradiation (T$_{irr}$), and a metallicity ratio (Z) expressed as a fraction of solar metallicity using values from \cite{asplund_chemical_2009}. The intrinsic temperature is derived from an incoming flux at the bottom of the atmosphere model, which serves as a boundary condition, using equation \ref{eq:T_int}:

\begin{equation}\label{eq:T_int}
      T_{int} =  \left( \frac{F_{int}}{\sigma_B}\right)^{\frac{1}{4}} \,
.\end{equation}
The expressions of T$_{irr}$ and T$_{int}$ are similar to those found in \cite{guillot_radiative_2010}.

\verb+Exo-REM+ can evaluate the gravity throughout the atmosphere using the hydrostatic balance equation (HSE). The gravity varies with pressure and altitude. It is hence possible to evaluate how gravity varies at the bottom of different atmospheres as a function of the model parameters, and most notably the stellar irradiation, as shown in Figure \ref{profiles_insolation}. \verb+Exo-REM+ is run from a pressure of 10$^{-4}$ to 10$^{4}$ bar. The list of molecules considered and the references for the line lists used are given in Table \ref{table:list chemicals}.
\footnote{Further details on Exo-Rem can be found here: https://gitlab.obspm.fr/Exoplanet-Atmospheres-LESIA/exorem and in \cite{blain_1d_2021}}

\begin{table}[h!]
\centering
\scalebox{0.8}{
\begin{tabular}{c c}
\toprule
Species & Line list \\
\midrule
CH$_4$ & TheoReTS (1)\\
CO   & HITEMP  (2)\\
CO$_2$   & HITEMP  (2)\\
FeH & ExoMol  (3)\\
H$_2$O   & HITEMP (2)\\
H$_2$S  & ExoMol (4)\\
HCN  & ExoMol (5)\\
K  & NIST (6)\\
Na  & NIST (6) \\
NH$_3$  & ExoMol (7,8) \\
PH$_3$  & ExoMol (9) \\
TiO  & ExoMol (10) \\
VO  & ExoMol (11) \\
 \hline
\end{tabular}}
\caption{References of the different sources of molecular opacities (1) \cite{rey_accurate_2017}; (2) \cite{rothman_hitemp_2010}; (3) \cite{bernath_mollist_2020}; (4) \cite{azzam_exomol_2016}; (5) \cite{harris_improved_2006}; (6) \cite{kramida_nists_2019}; (7) \cite{coles_exomol_2019}; (8) \cite{yurchenko_theoretical_2015}; (9) \cite{sousa-silva_vizier_2014}; (10) \cite{schwenke_opacity_1998}; (11) \cite{mckemmish_exomol_2016}}
\label{table:list chemicals}
\end{table}

\subsection{Interior model}\label{Interior model}

The interior model used in this work is \verb+Exoris+ \citep{licari_microscopic_2016,mazevet_ab_2019,mazevet_benchmarking_2022}, which solves the hydrostatic equations for a given set of ab initio equations of state. Two regions are considered: an H-He envelope and a core that can be divided into an outer ice and an inner rocky part. The envelope is considered adiabatic and consists of a mixture of hydrogen and helium at a given mass fraction denoted $\gamma_{He}$. The core of mass $M_c$ is considered isothermal. The isentrope is computed by interpolation in an EOS dataset at a constant entropy that is fixed at the outer boundary for a given temperature and pressure. The hydrostatic equations are solved for a number of layers of fixed composition and mass and lead to a radius corresponding to the mass of the planet considered.  The EOSs considered in this work for the envelope are the SCVH \citep{saumon_equation_1995}, which is used to reproduce the results of previous work, and the reevaluated CMS ab initio EOS \citep{mazevet_benchmarking_2022}, which is used to produce our results. The water ab initio EOS \citep{mazevet_ab_2019} is used to describe the core, as is the MgSiO3 EOS from \cite{mazevet_ab_2019}.

For both H-He EOSs, the additive rule for mixing hydrogen and helium is used. Recent works, such as those of \cite{chabrier_new_2021} and \cite{howard_accounting_2023}, have shown that corrections to the ideal mixing should be taken into account. Most notably, \cite{chabrier_new_2021} suggest that by adapting the hydrogen EOS, it is possible to take into account the interactions between hydrogen and helium. However, as pointed out in \cite{mazevet_benchmarking_2022}, such an ad hoc procedure incorporates the intrinsic differences arising from either the ab initio results or the fitting procedure used to produce the EOS. As this introduces an uncontrolled parameter into the evaluation of the H-He EOS, we willingly chose not to adopt this approach.

Research on the internal structure of Jupiter has experienced a surge in recent years, marked by significant contributions from missions such as Juno. By integrating insights from these missions with formation models, there has been a notable increase in the complexity of interior models
\citep{vazan_evolution_2016,venturini_planet_2016,wahl_comparing_2017,debras_new_2019,valletta_deposition_2019,ormel_how_2021-1,debras_superadiabaticity_2021,miguel_jupiters_2022,helled_revelations_2022,militzer_juno_2022,idini_gravitational_2022}. In order to fit the higher-order gravitational moments (above J2), models proposing inhomogeneous interiors with dilute cores and/or superadiabaticity are being proposed. We however chose to remain with a two-layered approach: a homogeneous envelope and a solid core, which reduces the number of uncontrolled parameters. This approach appears to be validated by the latest results of \cite{bloot_exoplanet_2023}. Through the comparison of a homogeneous and a non-homogeneous interior model applied to exoplanets, these authors show that the two are degenerate with the current observations. They also show that the differences in bulk metallicity are not substantial, and find little difference between the resultant estimations of the core mass in most cases. Given these findings, and the fact that the homogeneity of the interior is ill constrained with the current observations, we do not consider it in  the remainder of this work. It should, however, be kept in mind that by using a two-layer model, we are formally representing only one example within a range of possible interior models and are neglecting the possibility that exoplanet cores could be diluted, as observed in the case of Jupiter.

\subsection{Joining models}\label{Joining models}

We chose to link the atmosphere and the interior  in a self-consistent way. In order to do this, we link on pressure (P$_{link}$), temperature (T), average molecular mass ($\mu$), and gravity ($g$).

\begin{figure}
\centering
\includegraphics[width=\hsize]{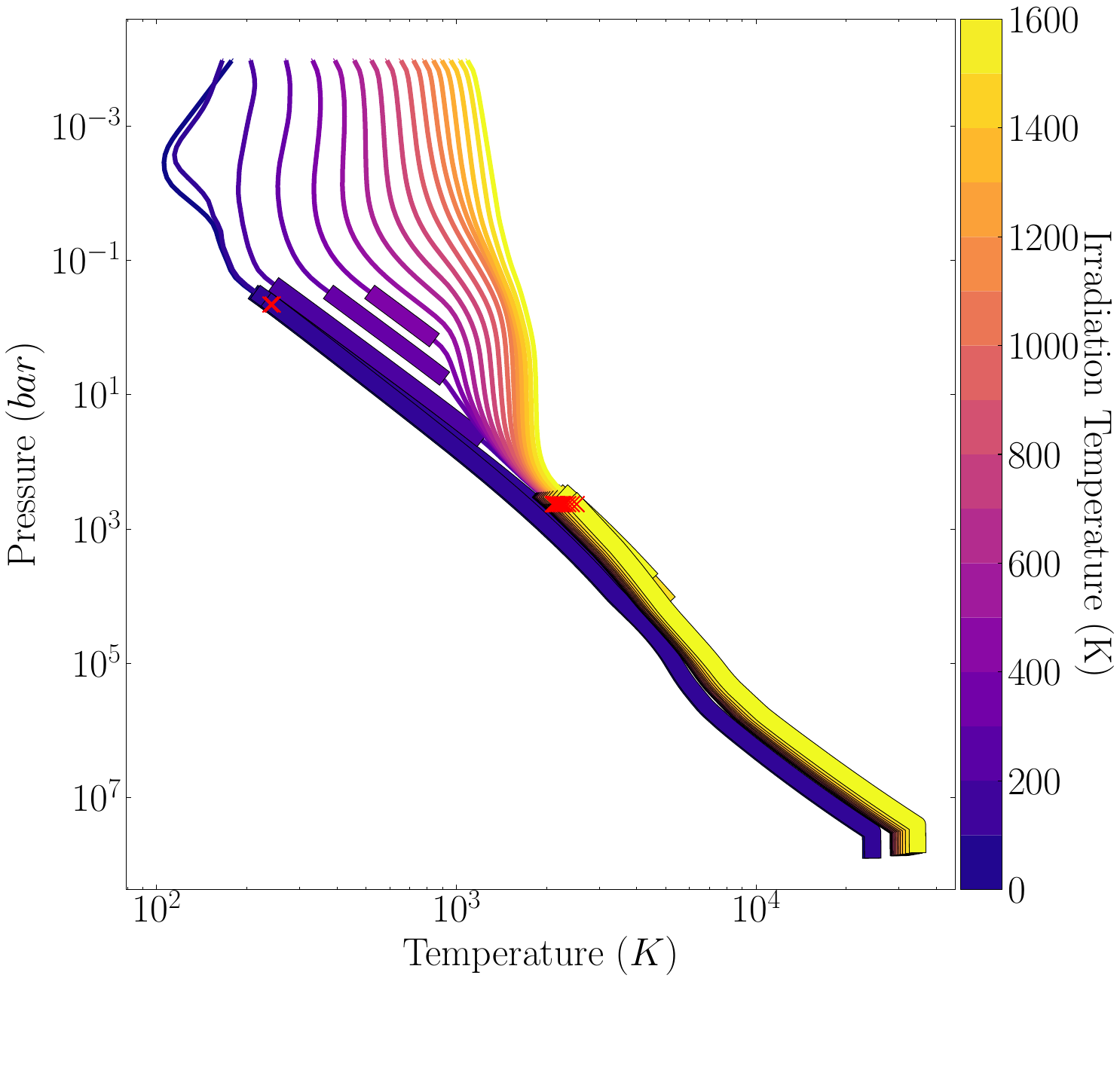}
\caption{Pressure--temperature profiles for a 1 M$_J$ planet across a range of stellar irradiation values. The red crosses indicate the point at which the interior model and atmosphere model are linked. The thicker zones delimited by black boxes correspond to the onset of convective dynamics in the atmosphere continuing down to the interior.}
     \label{profiles_insolation}
\end{figure}

The linkage pressure (P$_{link}$) is a function of the atmosphere's input properties. We can consider two regimes in the atmosphere, radiative and convective, which can be identified in Figure \ref{profiles_insolation} as the approximately isothermal regions for the former and the adiabatic regions for the latter. P$_{link}$  approximately corresponds to the radiative--convective boundary. In the cases where there ismore than one radiative region, P$_{link}$ is defined by the deepest radiative--convective boundary, at higher pressure. To correctly join the interior model to the atmosphere model, we need to ensure that we join uniquely within the convective region. As the interior model is calculated considering an isentropic profile, P$_{link}$ is in practice either equal to or at a slightly higher pressure than the radiative--convective boundary pressure. \verb+Exo-REM+ determines the convective regions of a given profile by checking that the temperature gradient is greater than the adiabatic gradient. Once a convective region is determined, we can then use the uppermost temperature and pressure values within the given region to run \verb+Exoris+. The deeper, most convective regions are depicted in Figure \ref{profiles_insolation} by the black boxed zones with crosses marking the linkage pressure. As the irradiation temperature decreases below the approximate value of the intrinsic temperature, we see that the atmosphere is convective at lower pressures. Once \verb+Exoris+ has computed a radius for this given pressure--temperature point at the chosen planetary mass, the gravity can be evaluated. As discussed in section \ref{Atmosphere model}, for \verb+Exo-REM+, the gravity at 1 bar is an input parameter and the gravity is evaluated using the HSE throughout the atmosphere. Therefore, the gravity value at 1 bar needs to be adjusted to ensure it matches the gravity at P${link}$, the top pressure point of \texttt{Exoris}, located deeper in the atmosphere. Consequently, \verb+Exo-REM+ and \verb+Exoris+ are run iteratively until the gravity at P${link}$ is matched.

The average molecular mass is linked in a direct way by adapting the mass ratio of helium:
\begin{equation}
      X_{H_2} + Y_{He} + Z_{Heavy} = 1 \,
.\end{equation}
Considering the interior mix as
\begin{eqnarray}
       X_{h_2} + (Y_{He})_{int} = 1, \\
      (Y_{He})_{int} = (Y_{He} + Z_{Heavy})_{atm} \,
,\end{eqnarray}
which can be linked to the average molecular mass given by the atmosphere at the P$_{link}$ using

\begin{equation}
      \frac{1}{<\mu>} = \sum_i \frac{X_i}{\mu_i} \,
.\end{equation}

The hydrogen fraction is derived from the atmospheric model when considering the mixing ratio of various species. Hence, heavier elements found in the atmosphere are also accounted for in the interior envelope. The current formulation paves the way for future improvements by including, for example, the EOS of water as a proxy for heavier elements.

For a converging model, only a few iterations are required, although this varies across the parameter space. The convergence criterion is set for an error of below 0.2$\%$ for each parameter.

To complete the analysis presented in this work, grids of individual models are constructed along the various physical dimensions (M, T$_{int}$, T$_{irr}$, Z, M$_c$). Individual models within the grid have different output values (e.g. radius and spectrum) along these dimensions but are built without altering these input parameters. Table \ref{tab:grid_layout} gives the layout of the grid, including the grid limits and the step used.

\begin{table*}[h!]
\caption{Explored parameter space }
\centering
\begin{tabular}{|c|c|c|}
 \hline
 Parameter & [min ; max] & Step\\
 \hline
 Mass M (Jupiter) range 1  & [0.1 ; 1] & 0.05 \\[0.1cm]
  Mass M (Jupiter) range 2  & [1 ; 10] & 0.5\\[0.1cm]
 Intrinsic Temperature T$_{int}$ (K) &  [0 ; 2000]  & 50\\[0.1cm]
Irradiation Temperature T$_{irr}$ (K) &  [0 ; 2000]  & 50\\[0.1cm]
 Metallicity Z (log10(X Solar)) & [0.01; 35] & log-spacing \\[0.1cm]
 Core M$_c$ (M$_E$)   & [1,40] & 3 \\[0.1cm]
 \hline
\end{tabular}
\label{tab:grid_layout}
\end{table*}

\subsection{Model comparison}
\label{Model comparison}

An overview of the features available in the models used here can be found in Table \ref{table:model comparaison}. \cite{marley_sonora_2021} is the first of a series of papers, and \cite{karalidi_sonora_2021} is the second paper in this series and adds disequilibrium chemistry. Clouds can be found in the latter \cite{morley_sonora_2024} model. The models are for the non-irradiated case; they are developed for emission spectra and do not provide transit spectra.

\begin{figure*}[t!]
\centering
   \includegraphics[width=17cm]{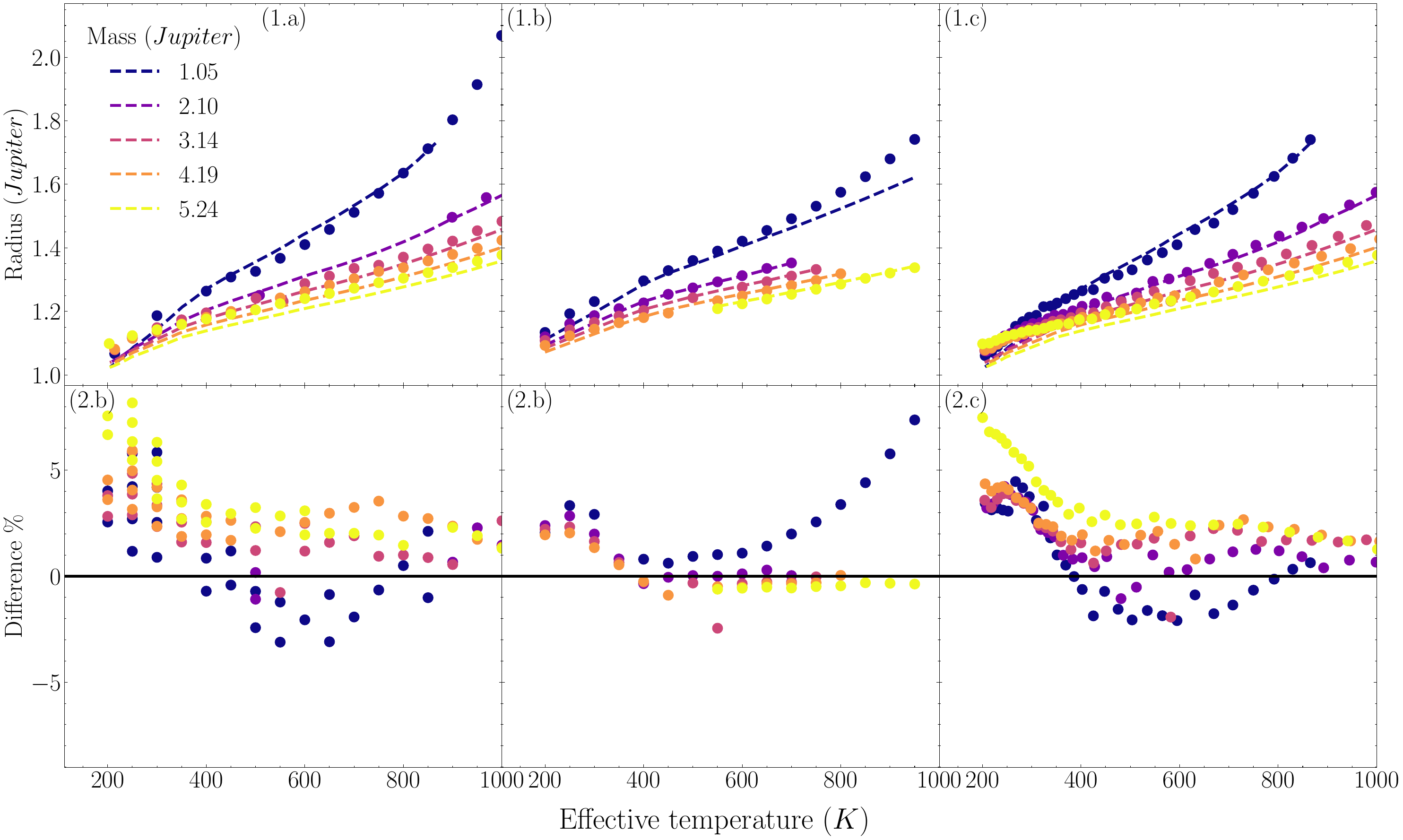}
     \caption{Comparison between results from this work and available work using similar physical properties. Top row (2):   (a) Comparison between the radius obtained in this work using the ab initio EOS of \citet{mazevet_benchmarking_2022} and the results of \cite{phillips_new_2020}, who use the ab initio EOS of \citet{chabrier_new_2019}.
     (b) Comparison between the radius obtained using the developed interior--atmosphere model using the SCVH EOS with the results of \cite{marley_sonora_2021}, who also use the SCVH EOS \citep{saumon_equation_1995}.
     (c) Our interpolated ab initio EOS grid compared to \cite{phillips_new_2020}.
     Bottom row (2a, b, and c): Corresponding percentage differences.}
     \label{Radius_comparaison}
\end{figure*}

Using the public evolution grid from \cite{phillips_new_2020}, as well as that from \cite{marley_sonora_2021}, we can compare the results given by our models for the non-irradiated case. Figure \ref{Radius_comparaison} illustrates a comparative analysis of planetary models within the 1-5 M$_J$ range. We conducted this comparison using both the ab initio EOS \citep{chabrier_new_2019} and the SCVH EOS \citep{saumon_equation_1995}, aiming to provide a comprehensive description of the variations between these models.

It is noteworthy that when directly comparing the calculations (non-interpolated), the disparities between the models and our findings remain modest, and below $5\%$ in most of the evaluated range. Interestingly, our results consistently exhibit slightly larger radii at 200 to 400K, regardless of the EOS used. Above 400K, our outcomes align closely with those of \cite{phillips_new_2020}. However, when considering lower masses, a noticeable divergence emerges between our findings and those of \cite{marley_sonora_2021} as temperatures increase. While differences in atmosphere models could contribute to the observed variations, it is important to acknowledge their relatively minor impact compared to the substantial discrepancies arising from distinct EOS selections. For the remainder of this work, we use the ab initio EOS for effective temperatures above 400K.

The generation of complete grids along each dimension presents a substantial computational and temporal burden. To address this, we propose leveraging multidimensional interpolation tools such as the linear barycentric interpolation function from the Python \verb+SciPy+ interpolation toolbox (LinearNDInterpolator). Careful consideration is essential when using interpolations within a grid, as inaccuracies at specific points or grid boundaries can disproportionately influence the resulting values.

In our verification process, we assess the reliability of the interpolation tool by comparing our interpolated results with those of \cite{phillips_new_2020}, using a dataset computed at masses distinct from those within our target grid. Remarkably, the interpolation yields highly comparable results and the rightmost graphs of Figure \ref{Radius_comparaison} are comparable to the equivalent graphs on the left, which use actual grid points. 

The above comparison shows the overall validity of our model compared to currently available models, this for the two different EOSs tested here. While this comparison is made for the non-irradiated case, there is no difference in the methodology used when adding stellar irradiation. The capacity to interpolate within the grid allows for more flexible use and reduced overall computation time when evaluating planetary parameters as done throughout the present work.

\begin{table*}[h!]
\centering
\scalebox{0.8}{
\begin{tabular}{|c|c|c|c|c|c|c|c|c|}
 \hline
 Model& Transit spectra & Emission spectra & Photometry & EOS & Disquilibrium chemistry & Metallicity & Clouds & C/O ratio\\
 \hline
 Current work   & \checkmark & \checkmark & \checkmark & Ab initio & \checkmark & [-2.0;1.4] & $\times$ & $\times$ \\[0.1cm]
  \hline
  \cite{morley_sonora_2024} & $\times$ & \checkmark & \checkmark & Ab initio & $\times$ & [-0.5;0.5] & \checkmark & $\times$ \\[0.1cm]
 \hline
  \cite{marley_sonora_2021} & $\times$ & \checkmark & \checkmark & SCVH & Added in \cite{karalidi_sonora_2021} & [-0.5;0.5] & $\times$ & [0.25;1.50] \\[0.1cm]
 \hline
   \cite{phillips_new_2020}   & $\times$ & \checkmark & \checkmark & Ab initio & \checkmark & 0 & $\times$ & $\times$ \\[0.1cm]
 \hline
   \cite{baraffe_evolutionary_2003}   & $\times$ & $\times$ & \checkmark &  SCVH & $\times$ & 0 & $\times$ & $\times$ \\[0.1cm]
 \hline
\end{tabular}}
\caption{Comparison of currently available and widely used evolution models.}
\label{table:model comparaison}
\end{table*}

\section{Stellar irradiation}\label{Stellar irradiation}

Considering stellar irradiation, we can define the effective tem- perature of a planet as

\begin{equation} \label{eq:T effective}
    T_{eff}^4 = (1-A_{Bond}) \cdot T_{irr}^4 + T_{int}^4\,
.\end{equation}

Planets cool over time by evacuating internal energy into space and reducing their entropy ($S$). This is summarised by the following equation,

\begin{eqnarray}\label{eq:Luminosity equation 1}
    \int_{M_{core}}^{M_{p}} T \frac{d S}{d t} dm = - 4 \pi \sigma R_{p}^{2} (T_{eff}^{4}-(1-A_{Bond}) \cdot T_{irr}^{4})
.\end{eqnarray}

By assuming an adiabatic (i.e. isentropic) envelope and by neglecting the contribution of the upper atmosphere (i.e. the radiative region) in the total entropy, the thermal evolution is given by

\begin{eqnarray}    
    \label{eq:Luminosity equation 2}
    \frac{dt}{dS} = \frac{\int_{r=0}^{r=R_{p}} T(r) \rho(r)r^2dr}{\sigma R_{p}^{2} T_{int}^{4}} \,
.\end{eqnarray}

For the remainder of this work, the Bond albedo ($A_{Bond}$) derived from \verb+Exo-REM+ is approximately zero, assuming an M star for the light source (BT-Settle spectrum at 3500K from \cite{allard_models_2012}) and a cloud-free atmosphere. As such, for irradiated planets, the equilibrium temperature is approximately equivalent to the irradiation temperature.

\begin{figure}
\centering
\includegraphics[width=\hsize]{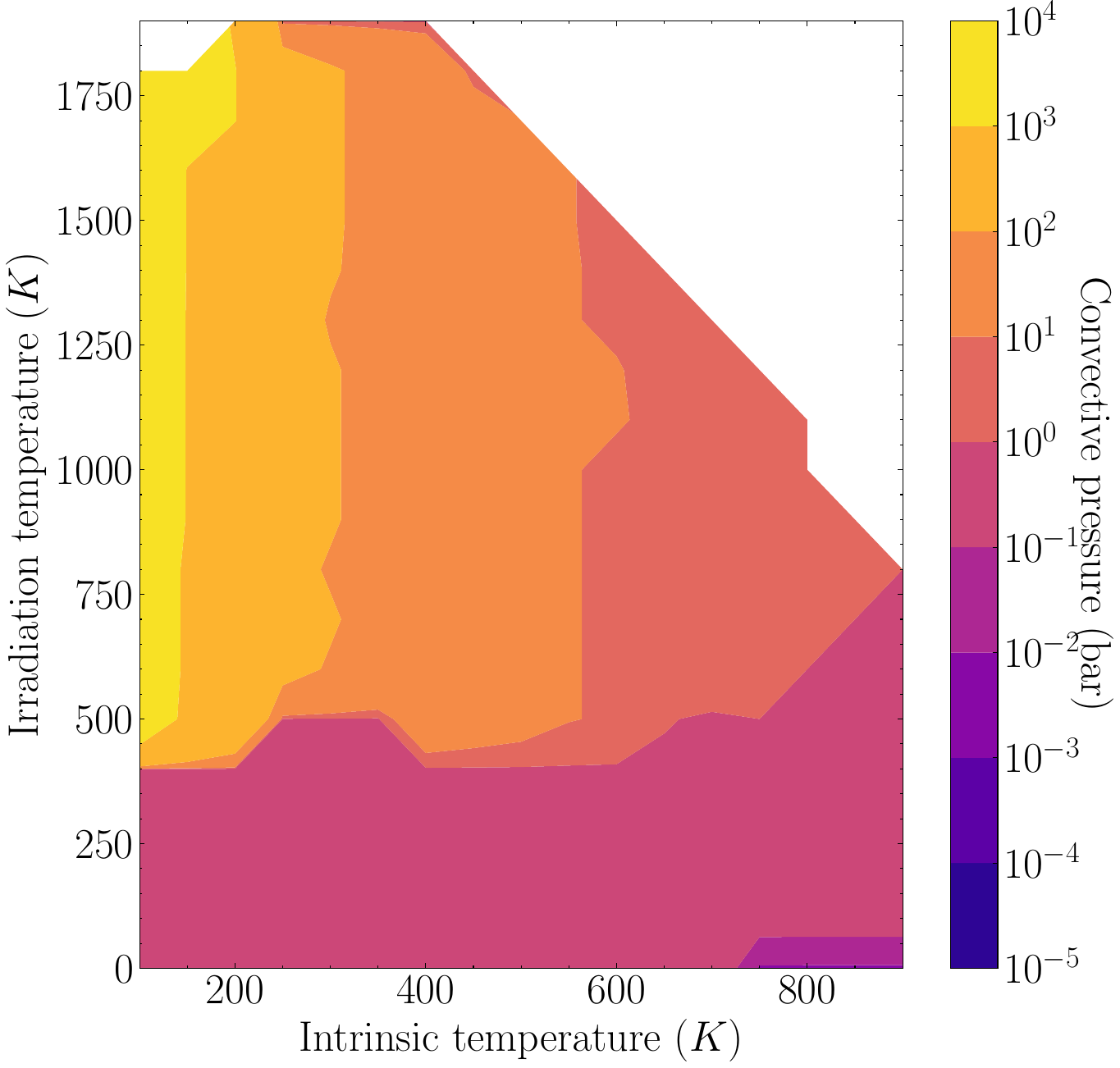}

  \caption{Schematical representations of the radiative-convective boundary (in pressure height) for varying intrinsic and irradiation temperatures for 1 M$_J$ planet.}
     \label{Convective_pressure}
\end{figure}

\begin{figure*}[t!]
\centering
   \includegraphics[width=17cm]{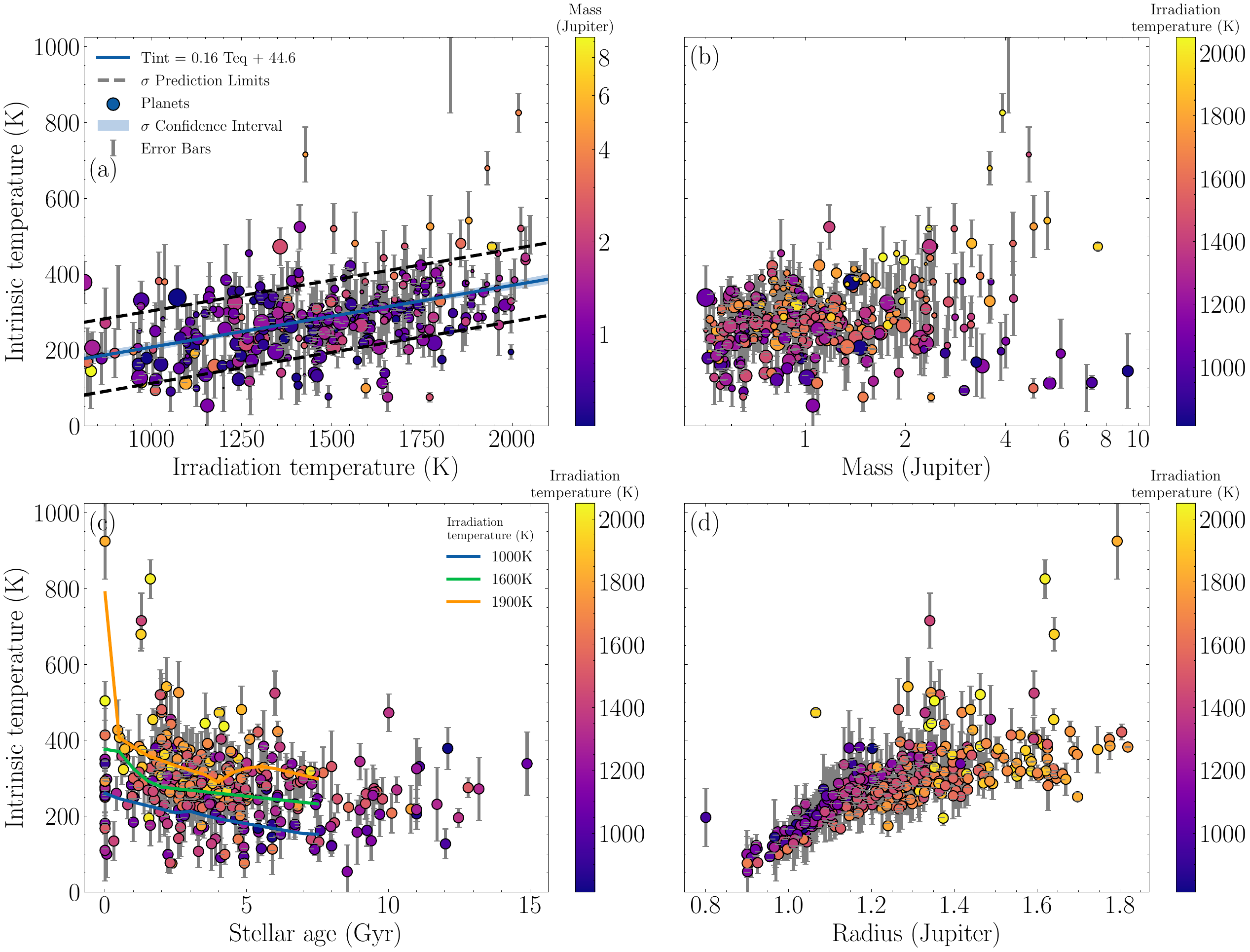}
     \caption{Intrinsic temperatures for planets between 0.5 and 10 Jupiter masses. (a) Intrinsic temperature with equilibrium temperature showing a linear increasing trend, with stellar age indicated qualitatively by point size. (b) Intrinsic temperature with mass, showing a greater spread at higher mass values, stellar age indicated qualitatively by point size. (c) Intrinsic temperature with stellar age, showing planets cooling with age, irradiated planets maintain a higher intrinsic temperature. Interpolated trends are shown for 1000 K, 1600 K, and 1900 K irradiation temperatures.  (d) Intrinsic temperature with radius showing the contribution to the radius from the irradiation. The list of planets used can be found in the appendix \ref{planet_list}.}
     \label{Planet_stats}
\end{figure*}

In the context of a given pressure--temperature profile for a planet, distinct radiative and convective regions become discernible. The location of the radiative-convective boundary, which demarcates these regions, hinges upon the equilibrium between stellar irradiation and the internal flux associated with planetary cooling. The result of this is shown in Figure \ref{profiles_insolation}, where an array of profiles 
is depicted, each characterised by a different irradiation temperature. Heightened irradiation extends the reach of the radiative domain to higher pressures within the atmosphere, a trend consistently revealed in Figure \ref{Convective_pressure} within the $($T$_{int};$T$_{irr})$ domain. This same figure shows how the radiative--convective equilibrium pressure (RCEP) evolves as a planet cools. For non-irradiated planets, there is little to no evolution of the RCEP with intrinsic temperature. Above approximately 400K in T$_{irr}$, there is significant burying of the RCEP for decreasing intrinsic temperatures, equivalent to a planet ageing. This result helps explain why non-irradiated models are valid for fixed linkage pressures between atmosphere and interior and why irradiated models need variability in this pressure. As the RCEP is driven to lower strata due to irradiation, the validity of certain assumptions regarding atmosphere models is put into question. Notably when the atmosphere model relies on the ideal gas law, significant deviation can arise at pressures above 10$^3$ bars. While we do not treat this in the present work, it remains important to ensure that the density does not deviate too greatly between the interior model and the atmosphere model. This can be the case for very old irradiated planets with T$_{int}< $150K, which Figure \ref{Planet_stats} shows to be a sparsely populated region considering current observations. Furthermore, models with fixed gravity throughout the atmosphere can suffer from deviations of up to 10$\%$ in gravity between the top and bottom of the atmosphere (1bar to RCEP). The adaptive linking implemented in this work accounts for this variability, as discussed in Section \ref{Joining models}.

As the heat penetrates further into the atmospheric layers, it notably elevates the entropy of the selected adiabatic gradient for the internal structure. Consequently, this upshifted gradient engenders a seemingly decelerated cooling rate for the planet. This is where a requirement emerges to establish an intricate relationship between age, mass, and radius within an irradiated evolution model for Jupiter-like exoplanets.

The self-coherent linking of the gravity between the atmosphere model and the interior model requires that we solve equation \ref{eq:Luminosity equation 1}  on an irregular grid in entropy. To complete this, equation \ref{eq:Luminosity equation 1} can be rewritten as equation \ref{eq:Luminosity equation 2} by considering a negligible contribution from the isothermal region of the core, where the entropy is variable. This is shown in \cite{mordasini_characterization_2012}; the luminosity of the  core is negligible during the evolution phase of a planet when compared to the internal luminosity. \cite{mordasini_characterization_2012} defines the internal luminosity as

\begin{equation} \label{eq:T effective}
    T_{int}^4 = \frac{L_{int}}{4\pi \sigma R^{2}}.\\
\end{equation}

We solve equation \ref{eq:Luminosity equation 2} using a Runge Kutta 4 numerical method along an interpolated regular grid in T$_{int}$ for a given set of planetary parameters. We use the specific entropies of young planets under the assumption of a hot start from \cite{marley_luminosity_2007} as initial entropies. A rigorous exploration of the impact of the initial entropy should be undertaken for the case of young giant planets whose ages are estimated to be below the Kelvin-Helmholtz timescale. Using equation 11 from \cite{guillot_giant_2015}, we find a timescale of $\approx 700 \text{ Myr}$ for 51 Eridani b using the best retrieval from section \ref{Case study : 51 Eridani b}.

In their study, \cite{thorngren_intrinsic_2019} established a connection between stellar irradiation and intrinsic temperature to explain the radius inflation of hot Jupiters by building upon their earlier research \citep{thorngren_bayesian_2018}. This earlier work established a link between the heating efficiency of incoming stellar radiation and its role in our understanding of the observed radius of a group of transiting planets. These authors used the \cite{thorngren_mass-metallicity_2016} mass--metallicity relationship to further increase the Bayesian heating efficiency fit. They propose a Gaussian fit between the model-based intrinsic temperature and the measured equilibrium temperature and radius. We chose to derive, via interpolation, intrinsic temperature values for planets within the NASA Exoplanet archive, taking into account the mass--metallicity trend identified in \cite{thorngren_mass-metallicity_2016}.

Figure \ref{Planet_stats} shows how the intrinsic temperature varies with observables. The intrinsic temperature values for each planet are obtained by interpolating from the grid of models. We identify a link between the intrinsic temperature and the equilibrium temperature. Figure \ref{Planet_stats}(a) shows that increased irradiation leads to an increased intrinsic temperature following a quasi-linear trend given by equation \ref{eq:Tint_trend}. 
\begin{equation}
\begin{array}{l}
T_{int} = 0.16\times T_{irr}+44.6
\end{array}
\label{eq:Tint_trend}
.\end{equation}

From a modeling perspective, this can be interpreted as a warming of the internal adiabat. Higher-mass planets exhibit a broader range of intrinsic temperatures, as shown in Figure \ref{Planet_stats}(b), likely due to their slower cooling rates and longer timescales for evacuating intrinsic energy compared to lower-mass planets. Figure \ref{Planet_stats}(c) shows that the intrinsic temperature that we derive is linked to stellar age. We would expect that a planet with a lower intrinsic temperature would be older. We observe that planets with higher irradiation temperatures maintain higher intrinsic temperatures. Finally, Figure \ref{Planet_stats}(d) shows that the intrinsic temperature increases for more inflated planets. There is a potential plateau that appears around the 400K region, where the radius is more sensitive to the intrinsic temperature. The stand-out samples far above the 400K intrinsic temperature region should be studied   further.

COmparing our results with those of \cite{thorngren_intrinsic_2019}, we observe no significant change in the trend of the intrinsic flux with increasing irradiation. In addition, the derived values for the intrinsic temperature are generally lower than those found by these authors. On the lower end of the irradiation temperature range, there is relatively good agreement between our results and theirs. However, as shown in Figure \ref{Planet_stats}(a), the majority of planets have a predicted intrinsic temperature of between 200 and 400K, which is as low as half of that shown in the fit by \cite{thorngren_intrinsic_2019}. This result corresponds to a heat transfer of $\approx0.09\%$ to the interior for a HD209458b-like planet \citep{henry_transiting_2000}, which is in agreement with the results of \cite{guillot_evolution_2002} for the flux transfer required to explain the radius of HD209458b, taking into account energy dissipation in the interior. Compared to the efficiency of heat transfer required in \citet{thorngren_intrinsic_2019}, this makes the observed inflated radii easier to reproduce by physical processes such as ohmic dissipation \citep{batygin_inflating_2010} or advection of heat by the atmospheric circulation \citep{tremblin_advection_2017}. We believe that equation \ref{eq:Tint_trend} should serve as an approximate temperature to be used when running global climate models for hot Jupiters, as is done for instance in \cite{komacek_patchy_2022}.

\section{Breaking degeneracies}\label{Breaking degeneracies}

To assess planetary parameters based on observable data, a commonly employed method is a statistical approach, such as Bayesian inference. Nevertheless, when faced with limited observables, it becomes intriguing to explore whether it is feasible to resolve parameter degeneracies using physical models. In this context, we focus on two specific parameters: metallicity and core size. Our objective is to investigate whether discrepancies in these parameters, within the (T$_{int}$; T$_{irr}$) space at a fixed mass, result in discernible variations in planetary radii.

\subsection{Metallicity}
By linking the molecular mass of the atmosphere in the convective region with the helium fraction of the interior, we can adjust the interior structure to take into account varying metallicities. We can then compare the $($T$_{int};$T$_{irr})$ space between different metallicities and uncover regions where the irradiation, age, and radius are sufficient variables to distinguish the planet's metallicity. 

\begin{figure*}[t!]
\centering
\includegraphics[width=\hsize]{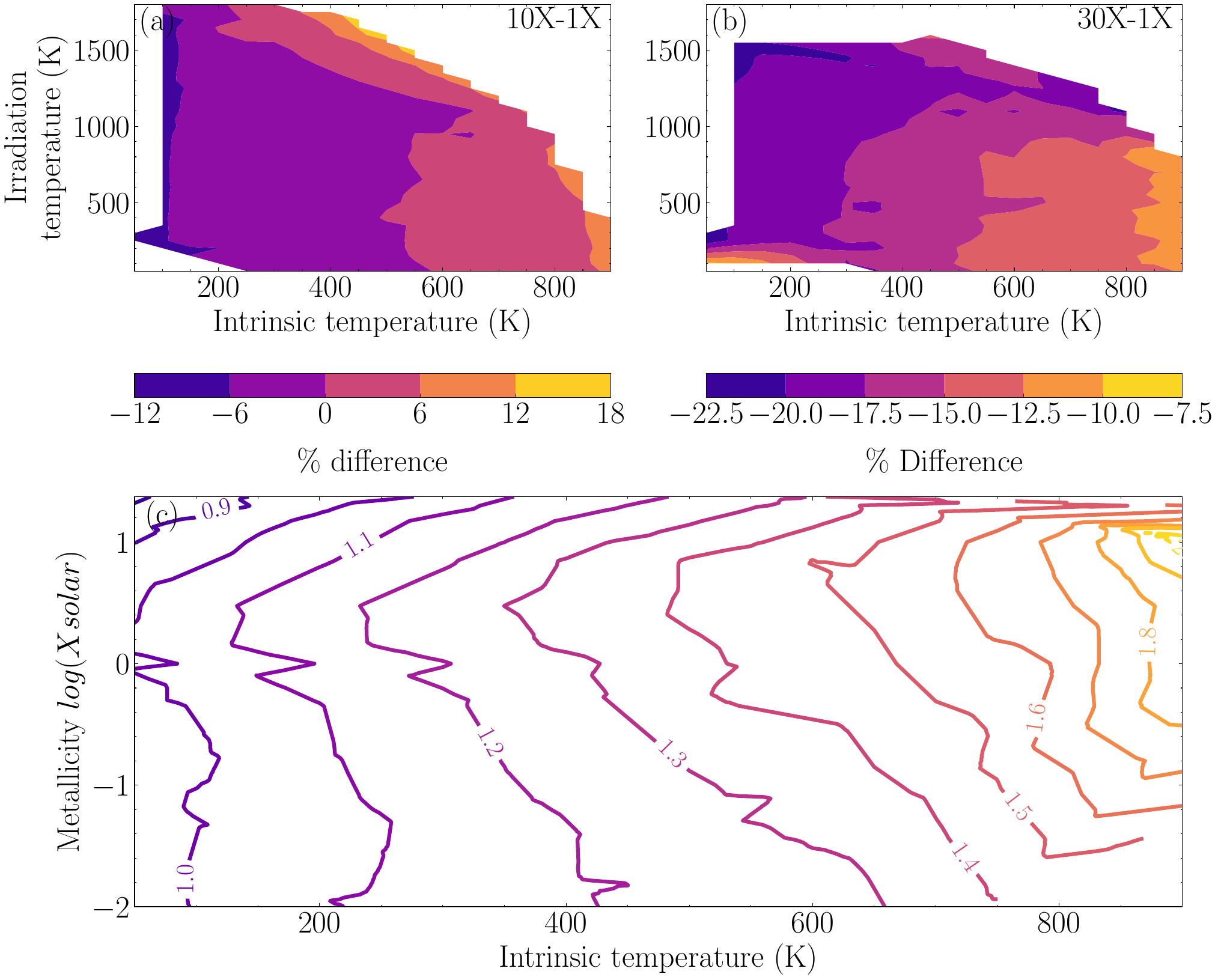}
  \caption{Radius difference as a percentage between a  1 M$_J$ planet with (a) 10 times or (b) 30 times solar metallicity  and an equivalent planet with 1 times solar metallicity. (c) Iso-radii lines in the metallicity--intrinsic temperature frame.}
     \label{Met_difference}
\end{figure*}

Figure \ref{Met_difference}(a) illustrates that a planet with ten times the solar metallicity and a very high intrinsic temperature may exhibit a larger radius than an equivalent planet with solar metallicity. In contrast, the right panel shows that a planet with 30 times the solar metallicity consistently maintains a smaller radius across the range of intrinsic and irradiation temperatures. The increase and subsequent decrease in the radius with metallicity is better shown in the lower panel of Figure \ref{Met_difference}. As we increase the temperature, the radius inflates for all cases, as expected. When increasing metallicity at fixed temperature, the radius also increases, but beyond a certain metallicity ($\approx$0.5) the radius decreases again. We therefore observe two regions, one where an increasing metallicity leads to an increase in absorption and hence the thermal effects dominate, and a second region where the molecular mass dominates and gravity forces the radius to contract. We therefore observe a `C' shape in the (Z;T$_{int}$) space. Thus, for a fixed radius, higher metallicities necessitate a lower intrinsic temperature because the radiative--convective boundary occurs at a lower pressure due to the increased opacity and greenhouse effect. However, beyond a certain metallicity threshold, the effect of the mean molecular mass on the atmospheric scale height dominates the effect of the opacity. Hence, higher intrinsic temperatures are required to maintain the radius and counteract the effects of gravity. The threshold is temperature dependent, 0.2 dex for 200K and 1 dex for 800K.

\subsection{Core size}

Exploring the properties of cores within gaseous planets is a topic of significant interest, and fundamental questions persist regarding their distribution within these planets and their contribution to both mass and radius. The layout of the core and even the presence of one, is a topic that has received significant attention where formation mechanisms are considered \citep{pollack_formation_1996, mordasini_characterization_2012}. However, when studying giant exoplanets, the primary question that arises pertains to whether variations in core characteristics have a discernible impact on observable parameters, such as the planet's radius. To explore this topic, we introduced an additional dimension to our analysis by considering different core sizes, from 1 to 40 times Earth mass (M$_E$).

\begin{figure}
\centering
\includegraphics[width=\hsize]{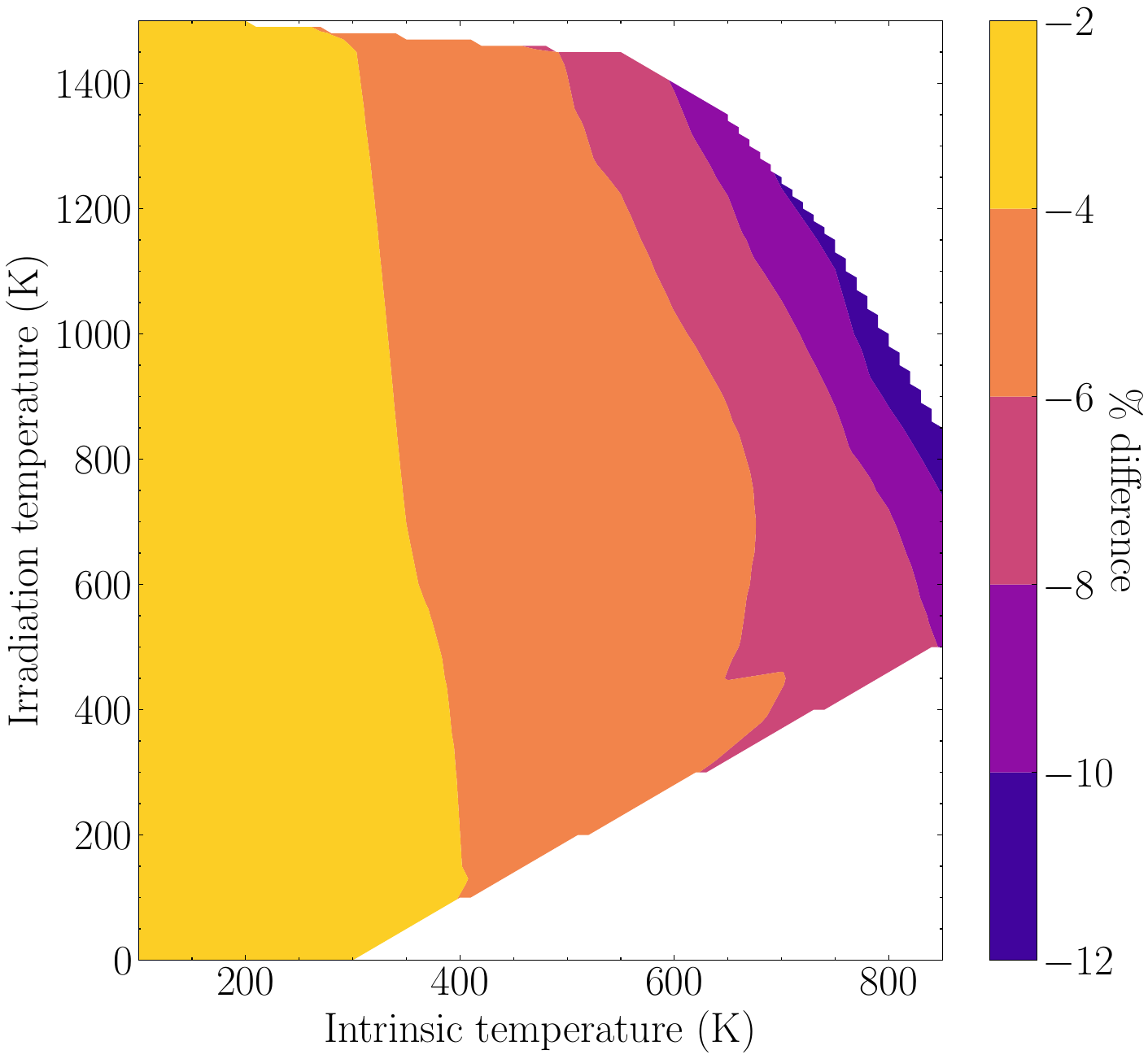}
  \caption{Radius difference as a percentage between a 20 M$_E$ core and 5 M$_E$ core.}
     \label{core_difference}
\end{figure}

In Figure \ref{core_difference}, we observe that at low intrinsic temperatures, the size of the core appears to be a degenerate parameter, showing minimal to negligible impact on the radius of a 1 M$_J$ planet. However, as we examine considerably younger and warmer planets with substantial radial extent, we notice an increasing influence of the core on the planet's radius. In fact, if a larger proportion of the planet's mass resides within the core, its ability to expand with rising temperature becomes constrained. This observation is supported by the slight tilt in the iso-radii differences.

Interestingly, we observe a weaker effect on the radius with increasing irradiation temperature compared to increasing intrinsic temperature. Consequently, the core of young planets has the potential to become a quantifiable parameter, so long as they have a sufficiently high intrinsic temperature. Exploration of this realm of the parameter space is poised to offer valuable insights into alternative interior designs, such as those featuring diluted cores.

\section{Grid interpolation and retrieval}\label{Grid interpolation and retrieval}
As discussed Section \ref{Model comparison} and shown in Figure \ref{Radius_comparaison} (c), it is possible to obtain the modelled radius of a planet at any point within the grid. Hence the radius is a continuous parameter. In order to efficiently develop grid retrieval tools, it is equally interesting to be able to interpolate the spectra (transit or emission) of a simulated planet with a given set of input parameters, defined as follows: 
\begin{eqnarray} \label{eq:planet}
    \vec{Planet} = (\mathrm{M}_p,\mathrm{T}_{int},\mathrm{T}_{irr},\mathrm{Z},\mathrm{M}_c)
.\end{eqnarray}

It is not possible to do this simply using the LinearNDInterpolator interpolator as for the radius. The issue is that this will lead to leakage between different wavelengths and the resulting spectrum will be heavily influenced by the distance between the old and new wavelengths, as well as the other planetary parameters. One solution is to create an array of interpolators, following: 
\begin{eqnarray}\label{eq:interp}
    f_{spectra}(\vec{Planet}) = [f_{\lambda_i}(\vec{Planet}) | \lambda_i \in \Lambda] \,
.\end{eqnarray}
For each wavelength of the generated spectra  ($\Lambda$), we create a unique interpolation function that takes the planetary parameters as input. It is no longer possible to leak information from one wavelength into another, as the unique interpolators are independent of one another. The drawback of this method is that it is essential to re-interpolate each point on the grid to a desired wavelength range before obtaining a new spectrum for a given planet.

Using Markov Chain Monte Carlo (MCMC) sampling on a continuous grid with the \verb+emcee+ Python library, we aim to ascertain the optimal planetary parameters that can elucidate an observed spectrum. To evaluate this approach, we introduce Gaussian noise ($\sigma=0.15\times<Flux>$) to a spectrum extracted from the grid, and therefore not interpolated.

The selected spectrum pertains to an artificial planet for which we fixed the parameters to 1 Jupiter mass, featuring intrinsic and irradiation temperatures of 800K and 100K, respectively, alongside solar metallicity and a 10 M$_E$ core. The resolution used is chosen so as to prepare for the test case in Section \ref{Case study : 51 Eridani b} ($R \approx 33$). By considering a H-band spectrum and irradiation temperature as observable factors, our assessment involves sampling the spectrum while treating the mass, intrinsic temperature, metallicity, and core mass as variables. We consider that the planet is placed at 10 parsecs from the observer. We adjust the flux level to account for this and the model radius using:

\begin{equation}
F_{observed} = F_{emitted} \cdot \frac{R_p^2}{D_{Earth-Planet}^2}
\label{eq:dilution}
.\end{equation}
The general expression of the likelihood function that we use in the MCMC is given as follows: 
\begin{equation}
\begin{array}{l}
\chi^2 = \sum_j^{n_{instr}}\frac{1}{n_j}\sum_i^{n_j} \frac{(f_{\lambda_{ji}} (Planet) - spectra_{\lambda_{ji}})^2}{\sigma_{spectra_{\lambda_{ji}}}^2} \\
+ \sum_k^{n_{obs}} w_k \frac{(Planet - obs_k)^2}{\sigma_{obs_k}}.
\end{array}
\label{eq:chi_2_eq}
\end{equation}
We consider two distinct terms, the first consisting of the double sum. This takes into account the fact that a spectrum can be a mosaic of spectra from multiple instruments. In this case, we propose to allow each instrument to be treated as an equally weighted observable. The second term is the simple sum that takes into account one or more physical characteristics of the planet, such as a mass measurement using radial velocity or a radius using the average transit depth. The proposed $w_k$ term is given to allow greater importance to be given to certain observed characteristics.

For the example studied here, equation \ref{eq:chi_2_eq} is reduced to the first term. We use uniform priors, except at the grid limits where we impose an infinite barrier. The priors are chosen as follows:

\begin{equation}
\mathrm{M}_p \in [0.5;3] \;|\; \mathrm{T}_{int} \in [500;1100] \;|\; Z \in [10^{-2};10] \;|\; \mathrm{M}_c \in [2;35]
\label{eq:priors}
.\end{equation}

\begin{figure}
\centering
\includegraphics[width=\hsize]{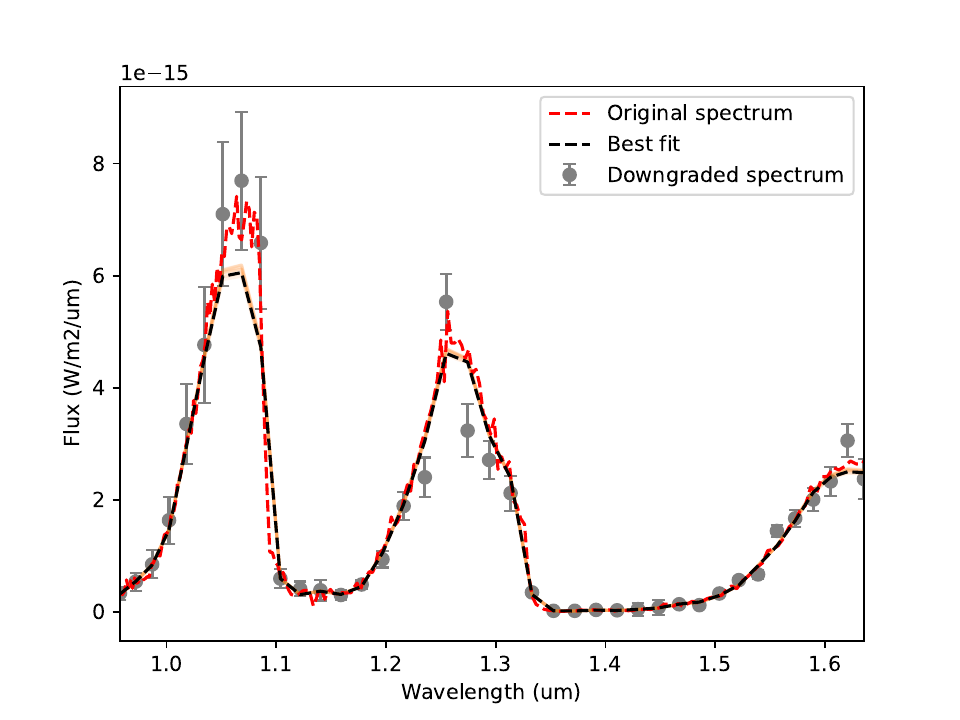 }
  \caption{Spectra of the overall best fit with 100 sample best fits over-plotted on an actual grid spectrum and a downgraded grid spectrum in H-band.}
     \label{test_fit}
\end{figure}

The outcome of this evaluation is illustrated in Appendix \ref{test_corner}, which presents a corner plot of the MCMC-based interpolated grid retrieval. The radius distribution is added by interpolating the four other parameters (M$_p$,T$_{int}$,Z,M$_c$) for each walker and iteration. Figure \ref{test_fit} shows the retrieved interpolated spectrum compared to the degraded and original spectra.

The results indicate a good capacity to retrieve a spectrum using interpolators and MCMC sampling; they reveal a relatively accurate retrieval of the intrinsic temperature, metallicity, and mass. The core mass is retrieved correctly, albeit with a relatively broad posterior distribution. This indicates that future core-mass estimations will at best be given with around a 5 M$_E$ uncertainty for a spectrum with similar characteristics to the one shown in Figure \ref{test_fit}.

\section{Case study: WASP-39 b, a transiting hot-Saturn}\label{Case study : WASP-39 b}
\label{Sec:WASP_39}

WASP-39 b is an inflated Saturn-like planet orbiting a G-type star with a period of P$\sim4 d$ \citep{faedi_WASP-39b_2011}. Radial velocity and transit photometry measurements as described by \citet{faedi_WASP-39b_2011} indicate a mass of $0.28 \pm 0.03$ M$_J$ and a radius of $1.27 \pm 0.04 $M$_J$, suggesting an inflated planet. \citet{ahrer_identification_2023} detected the presence of $\mathrm{CO_2}$ from its strong 4.4 $\mu m$ band from JWST NIRSpec-G395H observations, as part of the Early Release Science (ERS)  program. This detection suggests a relatively high metallicity ($\sim 10 \times$solar metallicity). We propose to use this JWST transit spectrum in an exemplary application of the model. Two approaches are adopted: in one instance, we assume no spectrum and uniquely mass and radius measurements, and hence equation \ref{eq:chi_2_eq} is reduced to the second term; while in another, we use the JWST ERS spectrum with the radial-velocity mass measurement. In this case, equation \ref{eq:chi_2_eq} includes the first term. The objective is to understand the impact that spectral information has on the constraint of parameters such as the intrinsic temperature derived for example in Figure \ref{Planet_stats}. We do not include a radius measurement as it is given within the transit depth at the JWST NIRSpec-G395H wavelengths. Radius measurements from, for example, \citet{faedi_WASP-39b_2011} correspond to the radius in the visible and near-infrared (R and Z bands).

\begin{figure*}[t!]
\centering
\includegraphics[width=14cm]{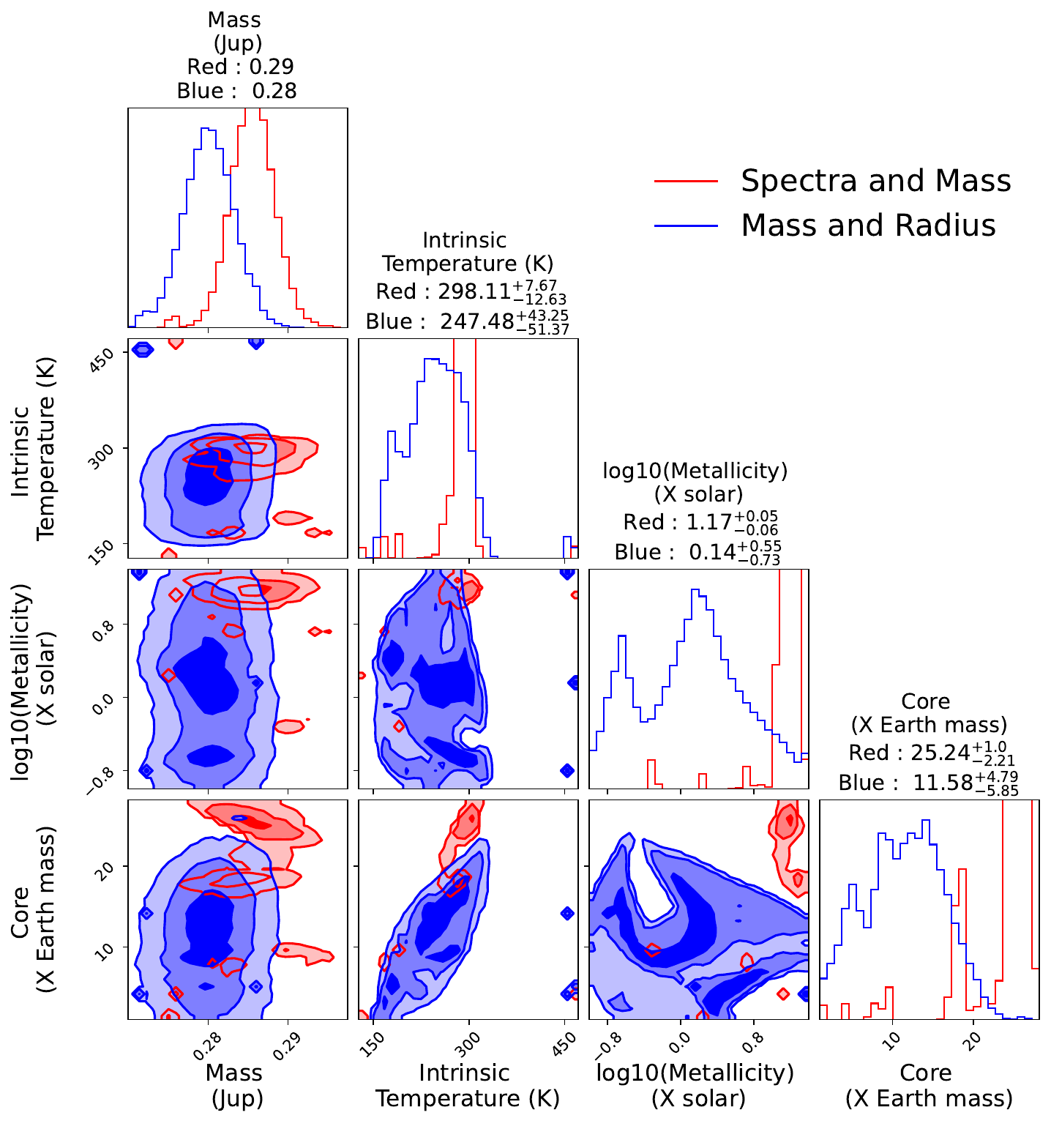}
  \caption{Posterior distributions of retrieval of the WASP-39 b JWST NIRSpec-G395H spectrum:  mass and radius (blue),  mass and spectrum (red).}
     \label{WASP39_corner}
\end{figure*}

\begin{figure}
\centering
\includegraphics[width=\hsize]{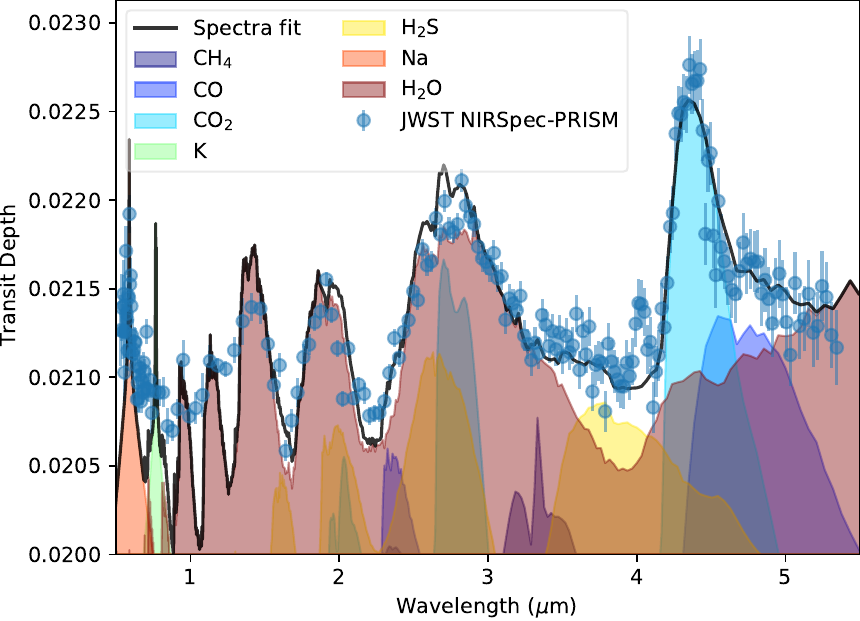 }
  \caption{Fit of the WASP-39 b JWST NIRSpec PRISM spectrum \citep{rustamkulov_early_2023} with contributions from the species considered. This fit was found for an intrinsic temperature of 297.39$_{-16.9}^{+8.95}$K and a metallicity of $1.16_{-0.09}^{+0.06}$dex.}
     \label{WASP39_fits}
\end{figure}

Figure \ref{WASP39_corner} shows the posterior distributions for the two scenarios. The retrieved values are given in Table \ref{table:Wasp39b}. We see that without the spectrum, the posterior distributions show a highly degenerated ensemble of potential parameters. Most notably, there are no constraints on the metallicity or the core mass. Possible values for the intrinsic temperature considering a Gaussian posterior with 1 $\sigma$ variation are between 150K and 320K. For the case with the spectrum, we ran the retrieval based on our grid of models using a chemistry at equilibrium. Stronger constraints are obtained for the metallicity and core mass. A relatively high metallicity of about ten times solar is necessary to adequately fit the $\mathrm{CO_2}$ band as shown in the best fit of Figure \ref{WASP39_fits}. A high T$_{int}$ of 297.39$_{-16.9}^{+8.95}$K is required to reproduce the inflated radius, given the equally high metallicity, but also to convert CH$_4$ into CO and CO$_2$ leading to the observed CO$_2$ spectral signature. The core proposed is relatively large, namely of between 20 and 30 times Earth's mass, and therefore represents up to 30$\%$ of the total planetary mass. To match the higher atmospheric metallicity, the helium mass fraction in the interior is approximately 0.50. As such, considering a solar-like helium mass fraction of 0.246 \citep{lodders_solar_2019}, the mass of heavy elements (core mass + excess helium) is approximately 45$\%$ of the total mass of the planet. Bloot et al. (2023) finds a core mass of 0.88 times Earth's mass using a homogeneous model and 0.90 times Earth's mass using a non-homogeneous core model. JWST data were not
used in either case. Furthermore, \cite{thorngren_connecting_2019} use a bulk metallicity of 22$\%$ of the planet's mass to derive an atmospheric metallicity of 40.51 times solar, again without using JWST data. We now know the atmospheric metallicity to be lower, and within the framework of their approach this could change how well mixed the planet is. In all cases, we find a core mass that is substantially higher with the JWST data. Without the JWST data, the result is very degenerate and would include both possibilities.
Figure \ref{WASP39_fits} shows the fit over-plotted on the NIRSpec-PRISM spectrum from \cite{rustamkulov_early_2023}. The $\mathrm{CO_2}$ signature is shown to be well fitted by the model. Discrepancies arise at shorter wavelengths due to the absence of clouds in the current model. The absence of photo-chemistry and $\mathrm{SO_2}$ in the model explains the missing absorption band around 4 $\mu$m. Nevertheless, this has no impact on the overall fit. However, we are able to reproduce the sodium signature at around 0.5 $\mu$m.   We conclude that adding spectroscopic measurements enables us to break the degeneracies between metallicity, intrinsic temperature, and core mass. The fact that the metallicity and core values found for the spectrum-constrained case are outside the 68$\%$ region of the unconstrained posterior distribution indicates that the spectrum provides essential information that is not included with mass and radius alone.
The metallicity derived here using the transit spectrum is in agreement with the approximately ten times solar value found by the extensive range of models applied to this planet and spectrum \citep{powell_sulfur_2024, constantinou_vira_2024,carone_wasp-39b_2023,feinstein_early_2023,the_jwst_transiting_exoplanet_community_early_release_science_team_identification_2023,alderson_early_2023}. Using the same spectrum and the retrieval code \texttt{TauREx} from \cite{ al-refaie_taurex_2021} ---and used for example in \cite{panek_re-analysis_2023}---, we find a metallicity ranging from 11.99 to 12.85 times solar. This result is also in agreement with the derived atmospheric metallicity found using the developed model. \cite{alderson_early_2023} find a wide range of intrinsic temperatures using three atmosphere  models: Atmo ($100$K), PHOENIX ($400$K), and PICASO+Virga ($100$K). This could be due to the lack of coupling between the atmosphere model and the interior model, to which particular attention was paid in the present study.

\begin{table*}[h!]
\centering
\begin{tabular}{|c||c||c|}
 \hline
 Parameter&Mass--radius retrieval&Mass--spectra retrieval\\
 \hline
 Mass (Jupiter)   & $0.29$ & $0.28$ \\[0.1cm]
 Intrinsic temperature (K) &  $250.80_{-52.37}^{+45.30}$ &  $297.39_{-16.9}^{+8.95}$ \\[0.1cm]
 Metallicity (log10($\times$ solar)) & $0.16_{-0.74}^{+0.58}$ & $1.16_{-0.09}^{+0.06}$ \\[0.1cm]
 Core (M$_E$)   & $11.43_{-5.96}^{+4.89}$ & $24.94_{-7.05}^{+1.20}$ \\[0.1cm]
 \hline
\end{tabular}
\caption{Best retrieval values for WASP-39 b considering the two performed retrievals}
\label{table:Wasp39b}
\end{table*}

\section{Case study: 51 Eridani b, a young  remote massive-Jupiter}\label{Case study : 51 Eridani b}
\label{Sec:51_Eri}
51 Eridani b (hereafter 51 Eri b) is a massive-Jupiter planet orbiting the young ($20 \pm 6 $ Myr) F0 type star \object{51 Eridani}. The high uncertainty on its age makes it difficult to interpret the photometric data in order to derive the planet's physical parameters. 51 Eridani b  was dentified in 2015 in high-contrast imaging on the basis of a $\mathrm{CH_4}$ absorption in its near-infrared spectrum \citep{macintosh_discovery_2015} using the Gemini Planet Imager \citep[GPI, ][]{macintosh_gemini_2014}. In a subsequent paper, \cite{rajan_characterizing_2017} used the same GPI Integral Field Spectrograph (IFS) data for 51 Eri b, finding an effective temperature ranging between 605 and 737 K, solar metallicity, and a surface gravity of log(g) = 3.5–4.0. \cite{samland_spectral_2017} show a supersolar metallicity at [Fe/H] = 1.03$_{-0.11}^{+0.10}$dex, a 759.51$_{-21.97}^{+21.34}K$ effective temperature, and a surface gravity of log(g) = 4.26$_{-0.25}^{+0.24}K$. More recently, \cite{brown-sevilla_revisiting_2023} used photometric data obtained with the Spectro-Polarimetic High-contrast imager for Exoplanets REsearch \citep[SPHERE, ][]{beuzit_sphere_2019} at the VLT to constrain  its  gravity to log(g)=$4.05 \pm 0.37$ and its radius to $0.93 \pm 0.04~R_J$. Unlike the previous case study described in Sect. \ref{Sec:WASP_39}, only spectrophotometric data are available. We can then reduce equation \ref{eq:chi_2_eq} down to the first term. This equally means that we have fewer observational constraints to infer the structure of the exoplanet.

Using the same observation as in \cite{brown-sevilla_revisiting_2023} and \cite{rajan_characterizing_2017}, we extracted a YJH spectrum using SPHERE/IFS \citep{claudi_sphere_2008} and the H spectrum using two GPI/IFS epochs. The observing conditions of the observations can be found in Table~\ref{tab:obs_table} of the Appendix. The extraction of a spectrum can be complicated and the uncertainties associated with this process must be kept in mind when subsequently running retrievals.

The pre-processing steps consist of corrections to the sky, the flat field, and the background, as well as bad pixel interpolation and wavelength calibration. These steps use the standard pipeline of the High Contrast Data Center, formerly known as the Sphere Data Center \citep{delorme_sphere_2017}. Regarding GPI, every raw frame was dark subtracted, flat-fielded, cleaned of correlated detector noise, bad pixel corrected, and flexure corrected using the standard GPI Data Reduction Pipeline (DRP; version 1.6) documented in \cite{perrin_gemini_2014, evans_gemini_2016}. The resulting set of frames was combined into a spectral datacube using Pyklip \citep{wang_pyklip_2015}, which allowed a more accurate wavelength and PSF recalibration in addition to frame centring.

As a post-processing algorithm, we consistently use \verb+PACO ASDI+ \citep{flasseur_paco_2020} 
for all datasets. \verb+PACO ASDI+  is an ADI-based algorithm that models the noise at a local scale in small patches. It relies on the learning of local temporal and spectral correlations of the noise to provide a reliable and statistically grounded signal-to-noise-ratio map and photometry confidence intervals. A full and detailed explanation of the algorithm can be found in \citet{flasseur_exoplanet_2018, flasseur_paco_2020, flasseur_robustness_2020} and its implementation on the Data Center and spectral optimization in \cite{chomez_preparing_2023}.

After measuring the contrast spectrum of the planet with PACO, that is the ratio of the flux coming from the planet to the stellar PSF flux, we obtained a physical spectrum by multiplying this contrast by a synthetic spectrum of the star based on BT-NextGen AGSS09 atmospheric models \citep{allard_models_2012}. The stellar spectrum is normalised by 2MASS \citep{skrutskie_two_2006} H magnitude. In addition to the H-band uncertainty, the final spectrum error budget also accounts for systematic uncertainties on the star PSF intensity \citep{wang_gemini_2014}, as well as for random uncertainties related to the PSF variations over the exposure and to PACO's fit. The two GPI spectra were finally averaged to get the H-band spectrum used in the present section. A similar procedure was applied to SPHERE data. Error bars for both SPHERE and GPI are displayed at 1$\sigma$.

\begin{figure}[t!]
\centering
\includegraphics[width=\hsize]{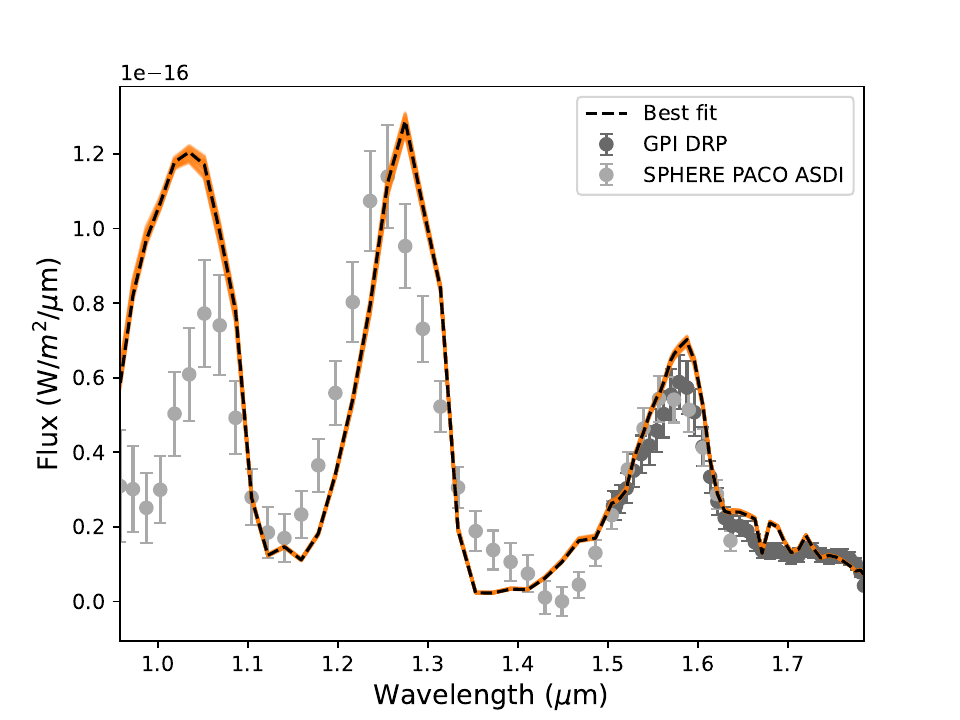 }
  \caption{Best fit of the observed spectrum of 51 Eridani b with 100 best fits.}
     \label{51ERIb_fits}
\end{figure}

Appendix \ref{51ERIb_corner} shows the corner plot of the spectral retrievals using MCMC, the best fit is shown in Figure \ref{51ERIb_fits}, and the retrieved values are given in Table \ref{table:51ERIb}. While this fit shows similar spectral features and an overall flux amplitude that is  close to the observed spectrum, there are notable differences below $1.1   \mu\text{m}$. These could be due to the presence of clouds, which are not included in the current version of the model. \cite{charnay_self-consistent_2018} show how clouds have an effect on emission spectra and most notably in the Y, J, and H bands. Appendix \ref{51ERIb_corner} indicates a multi-nomial distribution with three mass solutions found, one of which is further degenerate in intrinsic temperature and metallicity. All solutions are given in Table \ref{table:51ERIb}. In order to break the degeneracy, we include the age estimations found by solving equation \ref{eq:Luminosity equation 2}  in the corner plot. The three proposed ages are to be compared to the age estimate of the star of $20 \pm 6 \text{ Myr}$ proposed by \cite{macintosh_discovery_2015}. We identify the higher-mass case ($3.13_{0.04}^{0.05} M_J$) as being the most compatible configuration. This configuration has a corresponding age of $19.5 \pm 1.8$ Myr and is hence in strong agreement with the age of the system. It is however important to mention that the planet and the star need not be coeval. However, \cite{mordasini_characterization_2017} show that a 5 M$_J$ would take no longer than 10 Myr to finish its mass accretion with observations indicating less than 5 Myr. As such, when comparing the next closest age solutions (solutions 2 and 3) with the best match (solution 4), we see that the age of 19.5 Myr is a better match than 7 to 8 Myr.

As discussed in section \ref{Stellar irradiation}, the initial entropies used to evaluate the age are under the assumption of a  hot start. The MCMC posterior distribution shows a core mass of 31.86$_{-0.18}^{+0.32}$M$_{E}$ for the best matching age case. This result, within the models approximations, indicates the presence of a core and suggests a formation with core-accretion processes combined with the hot-start scenario. Such scenarios have been studied for example by \citet{berardo_hot_2017} and \citet{mordasini_characterization_2017}. \citet{mordasini_characterization_2017} show that it is possible for planets to undergo core accretion and have similar luminosities to those expected in both hot-start and warm-start scenarios. We explored the cold-start hypothesis for initial entropies, but all solutions found were below the estimated system age.

\begin{table*}[h!]
\centering
\begin{tabular}{|c||c|c|c|c|}
 \hline
 Parameter&Solution 1 & Solution 2 & Solution 3 & \textbf{Solution 4 (Best match)}\\
 \hline
 Mass (Jupiter)   & $1.53_{-0.04}^{+0.04}$ & $2.10_{-0.06}^{+0.02}$ & $2.16_{-0.09}^{+0.05}$ &  $\mathbf{3.13_{-0.04}^{+0.05}}$ \\[0.1cm]
 Intrinsic Temperature (K) &  $694.32_{-4.17}^{+5.17}$ &  $651.57_{-2.32}^{+1.65}$ &  $707.24_{-2.81}^{+2.83}$ &  $\mathbf{692.64_{-2.50}^{+2.52}}$ \\[0.1cm]
 Metallicity (log10($\times$ Solar)) & $-1.12_{-0.03}^{+0.03}$ & $-0.73_{-0.01}^{+0.01}$ & $-0.93_{-0.01}^{+0.01}$ & $\mathbf{-0.91_{-0.03}^{+0.03}}$ \\[0.1cm]
 Core (M$_E$)   & $4.52_{-0.28}^{+0.22}$ & $21.92_{-0.34}^{+1.27}$ & $34.88_{-0.16}^{+0.10}$ & $\mathbf{31.86_{-0.18}^{+0.32}}$ \\[0.1cm]
 Radius (Jupiter) & $1.39_{-0.01}^{+0.01}$ & $1.28_{-0.01}^{+0.01}$ & $1.27_{-0.00}^{+0.00}$ & $\mathbf{1.25_{-0.01}^{+0.01}}$ \\[0.1cm]
 log(g) & $3.31_{-0.02}^{+0.02}$ & $3.52_{-0.01}^{+0.01}$ & $3.54_{-0.02}^{+0.01}$ & $\mathbf{3.71_{-0.01}^{+0.01}}$ \\[0.1cm]
 log(age) & $6.23_{-0.03}^{+0.03}$ & $6.89_{-0.02}^{+0.02}$ & $6.86_{-0.02}^{+0.01}$ & $\mathbf{7.29_{-0.04}^{+0.04}}$ \\[0.1cm] 
 Age (Myr) & $1.70_{-0.11}^{+0.25}$ & $7.76_{-0.35}^{+0.37}$ & $7.24_{-0.33}^{+0.17}$ & $\mathbf{19.50_{-1.72}^{+1.88}}$  \\[0.1cm]
 \hline
\end{tabular}
\caption{Best retrieval values for 51 Eri b for all possible configurations: retrieval 4 corresponds to the best match considering the system's age. \tablefoot{The errors correspond to a 1 sigma deviation in the posterior distributions. Table \ref{tab:retrieval_capacity} based on Figure \ref{test_corner} shows more accurate retrieval errors. The systems age is estimated by \cite{macintosh_discovery_2015} to be $20 (\pm 6) \text{ Myr.}$}}
\label{table:51ERIb}
\end{table*}

The radius of 1.25R$_{J}$ found here is inconsistent with the `New nominal' radius found by \cite{brown-sevilla_revisiting_2023} of $0.93\pm0.04 R_J$. The radius is a parameter deduced from the input parameters as defined by equation \ref{eq:interp}. The model flux is scaled according to equation \ref{eq:dilution}. As such, the radius deduced through the model is directly linked to the flux scaling as well as the input parameters. This is not the case with purely atmospheric retrievals, where the radius becomes a free input parameter.

\section{Discussion}\label{Discussion}
The two examples show the advantages of a coherent model such as the one proposed here. The internal links between the mass, radius, metallicity, core size, and intrinsic temperature allow physically coherent results to be derived. Trends in parameters, such as intrinsic temperature derived from mass, radius, and effective temperature, need to be treated with utmost caution. The case of WASP-39b exemplifies this; albeit a planet with a mass below those studied in Figure \ref{Planet_stats}, it represents an interesting case where the metallicity and chemical composition derived from the spectrum drives the model towards a higher intrinsic temperature and helps explain the apparent inflated radius. WASP-39 b showed that deriving the intrinsic temperature as well as the internal structure from the mass and radius can lead to a wide range of possible chemical and thermal structures. Repeating this over the range of observed exoplanets can lead to increased uncertainties on overall trends. Using spectral data, it is possible to constrain the thermal and structural properties in a way that should pave the way for a re-evaluation of planetary statistics. \cite{thorngren_intrinsic_2019} use metallicity trends from \cite{thorngren_mass-metallicity_2016}, which provides a constraint that corresponds to having a spectrum. However, the uncertainty related to this metallicity trend is large. This leads us to believe that it is best to derive the metallicity from the spectrum on a planet-wise basis. This will become possible with the range of planets observed with JWST \citep{greene_characterizing_2019, beichman_direct_2019} and subsequently the dedicated Ariel mission \citep{tinetti_atmospheric_2020}.

The fits obtained for directly imaged planets need to rely on physical models that are self-consistent in order to correctly determine physical characteristics. Commonly used methods for decoupling the radius from the model will potentially lead to erroneous conclusions due to the contribution of nonphysical radii in adjusting the overall amplitude of the spectrum. The model proposed here has the potential to provide the most physically coherent parameters for directly imaged planets, even without prior knowledge of their age.

Four essential elements are missing from the current model, namely the C/O ratio, the H$_2$O EOS, clouds, and a diffuse core. The C/O ratio is famously hard to derive; it would be interesting to evaluate our capacity to constrain this parameter within a physically coherent model. The H$_2$O EOS is essential for exploring higher-metallicity planets, and especially Neptune and sub-Neptune-like planets, and might also have an impact on the distribution of heavy elements in exoplanet interiors should for instance metals be over-represented by the core mass. Clouds represent an important feature within the spectrum, changing the flux intensity between shorter and longer wavelengths and changing the thermal evolution of a planet. As we have shown, not including this parameter with a self-consistent model such as the one presented here can lead to physically incoherent results for other parameters. While it may be possible to account for differences in a parametric way, a self-coherent model would be in the spirit of the current work. Finally, incorporating diffused cores and examining constraints on the extent of the core in the context of an increasing amount of exoplanet data may provide better insights into this process. As pointed out in Section 1, diffuse cores provide better explanations for the Juno data than non-diffuse cores. While the impact on the radius is limited for older planets and those with lower intrinsic temperatures, warmer, younger planets might provide greater insight into the interior design. This addition to the model might also help us to better understand the distribution of heavier metals found within the core and the envelope, leading to a better understanding of the core mass.

\section{Conclusions}\label{Conclusions}

We present a combined atmosphere and interior model, HADES (Heat Atmosphere Density Evolution Model), which will serve as a powerful tool for proposing potential structural configurations of Jupiter-like planets. We linked the atmosphere and interior in a self-consistent manner by converging the physical parameters at the interface. We subsequently reevaluated the intrinsic temperatures corresponding to the internal fluxes of a range of exoplanets. This led us to the conclusion that there is reduced need for ohmic dissipation or other mechanisms that lead to higher intrinsic temperatures to explain the observed radius. We show that, by using a grid of models, we can use Bayesian inference to derive potential parameters for observed exoplanets. The case of WASP-39 b demonstrates the importance of using spectra to constrain parameters and hence reinforces the importance of atmospheric surveys such as the one anticipated as part of the Ariel mission. The case of 51 Eri b showcases the use of a physically coherent model to correctly scale the flux with radius even if the fit is of lower quality. Our application to 51 Eri b also demonstrates the ability of the interior--atmosphere model to infer formation processes when a system's age is available. Further work is required to complete the model, most notably by adding clouds.

\begin{acknowledgements}
We thank the anonymous reviewer for their valuable comments on the manuscript. We also thank Raphaël Galicher and Lucas Teinturier (LESIA, Observatoire de Paris, France) for their time saving advice on MCMC usage and expert opinions on results.\\

This project has received funding from the European Research Council (ERC) under the European Union's Horizon 2020 research and innovation programme (COBREX; grant agreement n° 885593)\\

This work was granted access to the HPC resources of MesoPSL financed by the Region Ile de France and the project Equip@Meso (reference ANR-10-EQPX-29-01) of the programme Investissements d’Avenir supervised by the Agence Nationale pour la Recherche.\\

This work was supported by CNES, focused on AIRS on Ariel, and by the Programme National de Planétologie (PNP) of CNRS/INSU, co-funded by CNES.

SPHERE is an instrument designed and built by a consortium consisting of IPAG (Grenoble, France), MPIA (Heidelberg, Germany), LAM (Marseille, France), LESIA (Paris, France), Laboratoire Lagrange (Nice, France), INAF - Osservatorio di Padova (Italy), Observatoire de Genève (Switzerland), ETH Zürich (Switzerland), NOVA (Netherlands), ONERA (France) and ASTRON (Netherlands) in collaboration with ESO. SPHERE was funded by ESO, with additional contributions from CNRS (France), MPIA (Germany), INAF (Italy), FINES (Switzerland) and NOVA (Netherlands). SPHERE also received funding from the European Commission Sixth and Seventh Framework Programmes as part of the Optical Infrared Coordination Network for Astronomy (OPTICON) under grant number RII3-Ct-2004-001566 for FP6 (2004-2008), grant number 226604 for FP7 (2009-2012) and grant number 312430 for FP7 (2013-2016).\\

This work has made use of the SPHERE Data Centre, jointly operated by OSUG/IPAG (Grenoble), PYTHEAS/LAM/CeSAM (Marseille), OCA/Lagrange (Nice), Observatoire de Paris/LESIA (Paris), and Observatoire de Lyon (OSUL/CRAL).\\

This research has made use of the NASA Exoplanet Archive, which is operated by the California Institute of Technology, under contract with the National Aeronautics and Space Administration under the Exoplanet Exploration Program.

\end{acknowledgements}

%
%

\bibliographystyle{aa}
\bibliography{bibliography}
\begin{appendix} 
\onecolumn

\section{List of planets used}
\begin{figure}[h!]
\includegraphics[
  height=22cm,
  keepaspectratio,]{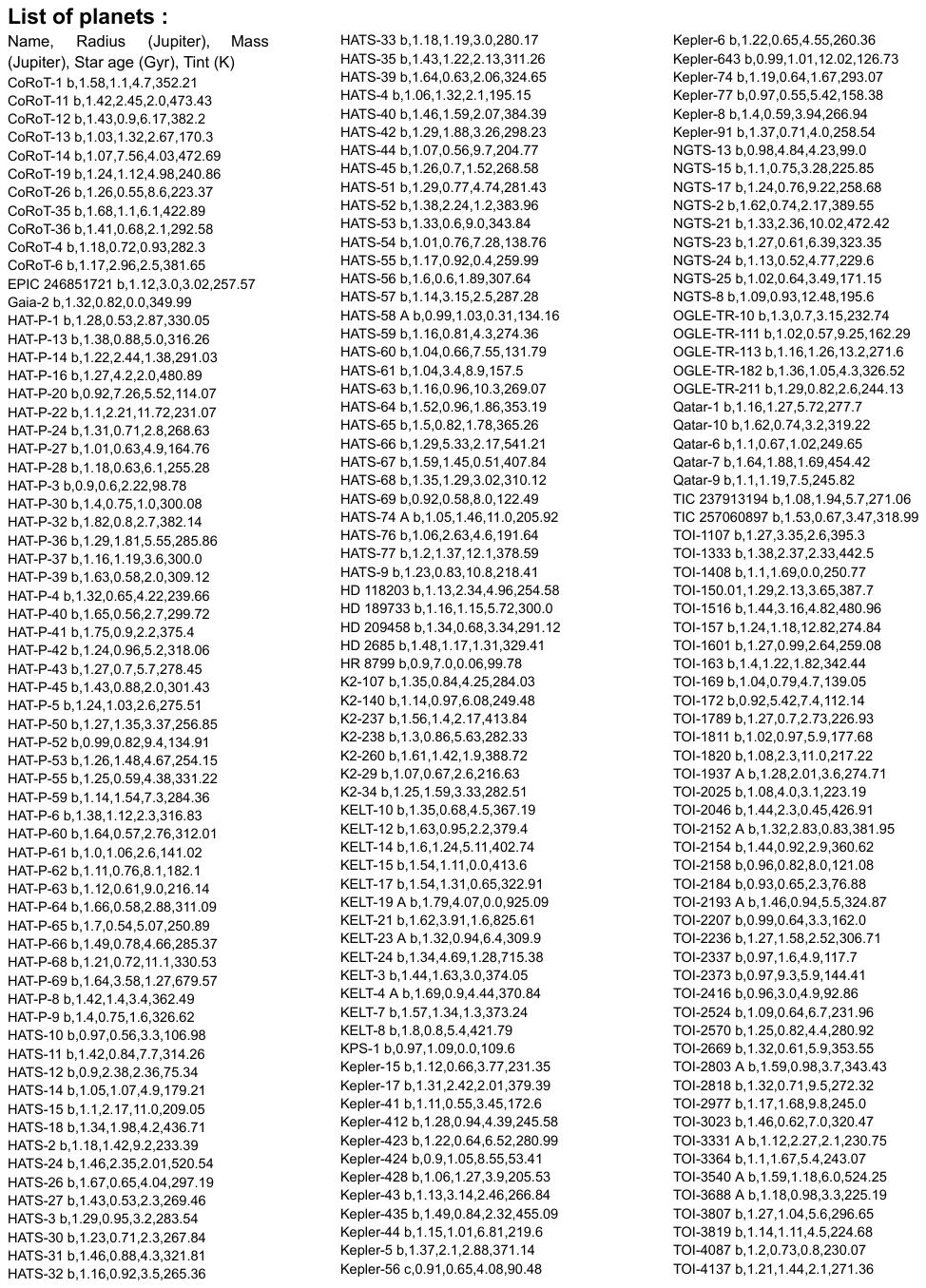 }
\caption{List of planets used to derive trends.}
\label{planet_list}
\end{figure}

\section{List of planets used}
\begin{figure}[h!]
\includegraphics[
  height=22cm,
  keepaspectratio,]{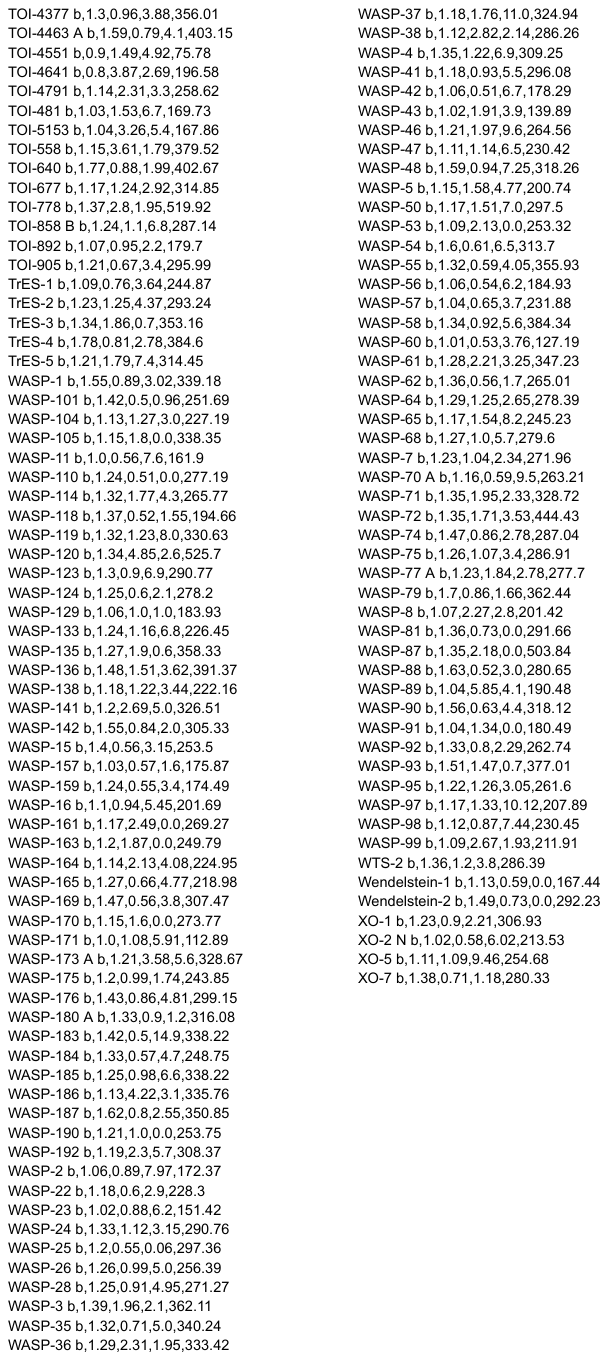 }
\label{planet_list b}
\end{figure}

\newpage
\section{Test MCMC result}
\begin{figure}[h!]
\includegraphics[
  height=17cm,
  keepaspectratio,]{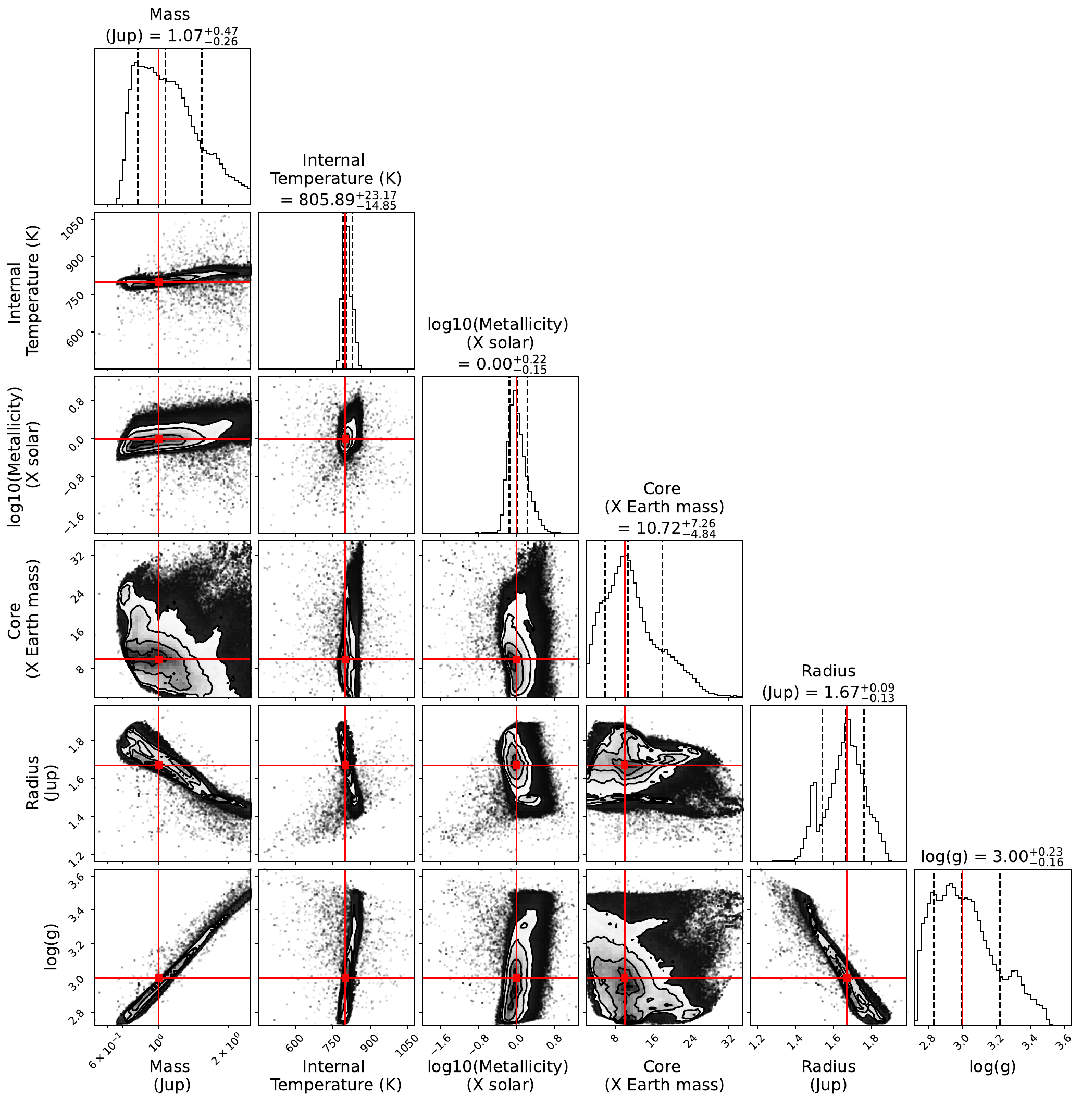 }
\caption{Corner plot of posterior distributions explaining downgraded grid test spectrum. Red cross-hairs indicate location of initial parameters of test spectrum.}
\label{test_corner}
\end{figure}

\begin{table}[h!]
\centering
\begin{tabular}{|c||c|c|c|}
 \hline
 Parameter&Original spectra&Retrieved Spectra&Difference ($\%$)\\
 \hline
 Mass (Jupiter)   & $1.00$ & $1.07_{-0.26}^{+0.47}$ &   $\approx 7.00\%$\\[0.1cm]
 Intrinsic Temperature (K) &  $800.00$  & $805.89_{-14.85}^{+23.17}$ & $\approx 0.74\%$\\[0.1cm]
 Metallicity (log10(X Solar)) & $0.00$ & $0.00_{-0.15}^{+0.22}$ &  $\approx 0.00\%$\\[0.1cm]
 Core (M$_E$)   & $10.00$ & $10.72_{-4.84}^{+7.26}$ &  $\approx 7.20\%$\\[0.1cm]
 Radius (Jupiter) &   $1.67$  & $1.67_{-0.13}^{+0.09}$ & $\approx 0.00\%$\\[0.1cm]
 log(g) & 3.00  & $3.00_{-0.16}^{+0.23}$ & $\approx 0.00\%$\\
 \hline
\end{tabular}
\caption{Comparison of the original spectrum's parameters with the retrieved parameters of the downgraded spectrum.}
\label{tab:retrieval_capacity}
\end{table}


\section{51 Eridani observing logs}
\begin{table*}[h!]
\centering
\begin{tabular}{l|c|c|c|} 
Instrument & SPHERE & \multicolumn{2}{|c|}{GPI} \\ 
\hline
OBS NIGHT & 2017-09-27& 2014-12-18 & 2018-11-20\\ 
FILTER & OBS$\_$H & GPI H filter & GPI H filter\\ 
DIT$\times$NDIT$\times$NEXP$^{~(a)}$ & 32$\times$1$\times$ 20& 59.6$\times$ 1$\times$38 & 59.6$\times$ 1$\times$60 \\ 
$\Delta$PA (°)$^{~(b)}$ & 51.8& 23.8 & 32.9\\ 
Seeing (")$^{~(c)}$ & 0.47 & ... & ...\\ 
Airmass$^{~(c)}$ & 1.09 & 1.14 & 1.16 \\ 
$\tau_0$ (ms)$^{~(c,d)}$ & 8 &...  &... \\ 
PROG ID & 198.C-0209(J)& GS-2014B-Q-500 & GS-2017B-Q-501 
\end{tabular} 
\caption{Observing log for the IFS observation of 51 Eridani used in this work.}
\tablefoot{ $^{(a)}$: DIT corresponds to the Detector Integration Time per frame, NDIT the number of DITs per exposure and NEXP the number of exposures in the template. $^{(b)}$: $\Delta$PA is the amplitude of the parallactic rotation. $^{(c)}$: values extracted from the updated Differential Image Motion Monitor (DIMM\footnote{\href{https://archive.eso.org/wdb/wdb/asm/dimm_paranal/form}{https://archive.eso.org/wdb/wdb/asm/dimm\_paranal/form}}) info, averaged over the sequence. $^{(d)}$: $\tau_0$ correspond to the atmosphere coherence time.}
\label{tab:obs_table}
\end{table*}

\newpage
\section{51 Eridani b MCMC result}
\begin{figure}[h!]
\centering
\includegraphics[width=\linewidth]{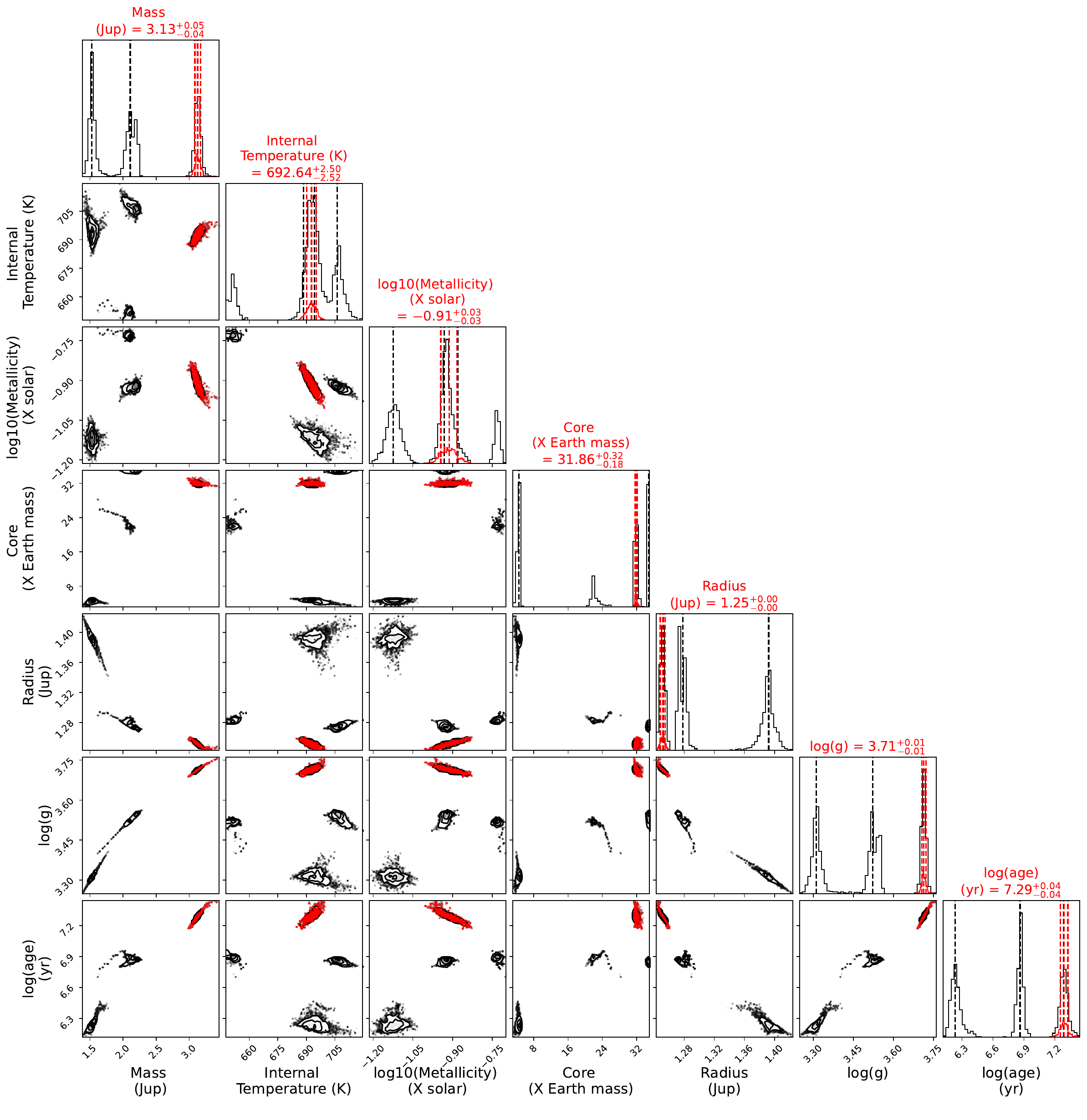}
\caption{Corner plot with posterior distributions of MCMC retrieval of the spectra of 51 Eridani b. Radius, log(g) and log(age) are derived parameters from the model. The red colored distribution indicates the most compatible parameters with the systems age.}
\label{51ERIb_corner}
\end{figure}


\end{appendix}

\end{document}